\documentclass[fleqn,usenatbib]{mnras}

\usepackage{newtxtext,newtxmath}
\usepackage[T1]{fontenc}
\usepackage{ae,aecompl}
\usepackage{blindtext}
\usepackage{adjustbox}
\usepackage{makecell}
\usepackage{enumerate}
\usepackage[switch,displaymath,mathlines]{lineno}
\usepackage{graphicx}   % Including figure files
\usepackage{amsmath}    % Advanced maths commands
\title[Mass loss in simple-population globular clusters]{Mass loss law for red giant stars in simple population globular clusters}

\author[Tailo, M. et al.]{
M. Tailo$^{1}$\thanks{E-mail:mrctailo@gmail.com, marco.tailo@unipd.it},  
A.\,P.\,Milone$^{1,2}$,
E.\,P.\,Lagioia$^{1,2}$,
F.\,D'Antona$^{3}$,
S.\,Jang$^{1}$,
E.\,Vesperini$^{4}$,
\newauthor
A.\,F.\,Marino$^{5}$,
P.\,Ventura$^{3}$,
V.\,Caloi$^{6}$,
M.\,Carlos$^{1}$,
G.\,Cordoni$^{1}$,
E.\,Dondoglio$^{1}$,
\newauthor
A. Mohandasan$^{1}$,
J.\,E.\,Nastasio$^{1}$,
M.\,V.\,Legnardi$^{1}$
\\
$^{1}$Dipartimento di Fisica e Astronomia ``Galileo Galilei'', Univ. di Padova, Vicolo dell'Osservatorio 3, Padova, IT-35122\\
$^{2}$Istituto Nazionale di Astrofisica - Osservatorio Astronomico di Padova, Vicolo dell'Osservatorio 5, Padova, IT-35122\\
$^{3}$Istituto Nazionale di Astrofisica - Osservatorio Astronomico di Roma, Via Frascati 33, I-00040 Monteporzio Catone, Roma, Italy\\
$^{4}$Department of Astronomy, Indiana University, Bloomington, IN 47405, USA\\
$^{5}$Istituto Nazionale di Astrofisica - Osservatorio Astrofisico di Arcetri, Largo Enrico Fermi 5, Firenze, I - 50125 \\
$^{6}$INAF -- IASF Roma, Via Fosso del Cavaliere, Roma, Italy, IT-00133  \\
}

\date{Accepted 2021 February 22. Received 2021 February 22; in original form 2020 December 30}

\pubyear{xxx xxx}

\begin{document}
\label{firstpage}
\pagerange{\pageref{firstpage}--\pageref{lastpage}}
% \linenumbers
\maketitle
% Abstract of the paper
\begin{abstract}
The amount of mass lost by stars during the red-giant branch (RGB) phase is one of the main parameters to understand and correctly model the late stages of stellar evolution.
Nevertheless, a fully-comprehensive knowledge of the RGB mass loss is still missing.
 
 Galactic Globular Clusters (GCs) are ideal targets to derive empirical formulations of mass loss, but the presence of multiple populations with different chemical compositions has been a major challenge to constrain stellar masses and RGB mass losses. Recent work has disentangled the distinct stellar populations along the RGB and the horizontal branch (HB) of 46 GCs, thus providing the possibility to estimate the RGB mass loss of each stellar population. The mass losses inferred for the stellar populations with pristine chemical composition (called first-generation or 1G stars) tightly correlate with cluster metallicity.  This finding allows us to derive an empirical RGB mass-loss law for 1G stars.  
 
 In this paper we investigate seven GCs with no evidence of multiple populations and derive the RGB mass loss by means of high-precision {\it Hubble-Space Telescope} photometry and accurate synthetic photometry. 
 We find a cluster-to-cluster variation in the mass loss ranging from $\sim$0.1 to $\sim$0.3 $M_{\odot}$.
 The RGB mass loss of simple-population GCs correlates with the metallicity of the host cluster. The discovery that simple-population GCs and 1G stars of multiple population GCs follow similar mass-loss vs.\,metallicity relations suggests that the resulting mass-loss law is a standard outcome of stellar evolution.  
\end{abstract}

% Select between one and six entries from the list of approved keywords.
% Don't make up new ones.
\begin{keywords}
stars: evolution, (stars:) Hertzsprung-Russell and colour-magnitude diagrams, stars: horizontal branch,  stars: low-mass Stars, stars: mass-loss,(Galaxy:) globular clusters: general
\end{keywords}

\section{Introduction}
\label{sec:into}

 A proper understanding of the late stages of stellar evolution depends on the precise knowledge of the law ruling mass loss along the red giant branch (RGB). Hence, determining the RGB mass loss law is a crucial step to fully understand stellar evolution.

To date, we still lack a conclusive theoretical description of RGB mass loss, and we mostly rely on empirical determinations. 
Historically, the law by \citet{reimers_1975} based on Population\,I stars has represented for decades the state of the art for describing RGB mass loss \citep[see also][and references therein]{fusipecci_1978, catelan_2000}. 
More recently, new formulations, based on magneto-hydrodynamics, have been proposed \citep[see][]{Schroder_2005,schroder_2007,Cranmer_2011} either as new law or as modifications of the \cite{reimers_1975} one. These new formulations tie mass loss to the interactions between surface turbulence and the magnetic field of the stars which are most relevant in the last part of the RGB where, indeed, most of the mass loss is predicted to take place.

A constantly updated observational framework is therefore crucial in the calibration and construction of the theoretical framework. \citet[][]{origlia_2007, origlia_2014} estimated the mass loss of 47 Tucanae and other fourteen GCs based on the excess of mid-IR light and suggested that a fraction of stars can lose mass at any RGB luminosity. To do this, they exploited  multi-band photometry from the Spitzer space telescope and from NIR ground-based facilities 
\citep[but see ][for a different interpretation of the photometric results by Origlia and collaborators]{boyer_2010,McDonald_2011}. \citet{momany2012a} on the other hand did not detect any NIR excess among RGB stars of 47\,Tucanae.
Mass loss has been also estimated by using Spitzer Infrared Spectrograph spectra of RGB stars, a technique adopted by \cite{McDonald_2011b} to investigate the brightest RGB stars in $\omega$ Centauri.

The comparison between the stellar mass of horizontal branch (HB) and RGB stars may provide an efficient approach to infer the RGB mass loss in a simple stellar population. 
Indeed, after reaching their tip luminosity, RGB stars undergo the so-called helium flash, namely an abrupt ignition of their degenerate helium core. After  this violent process, HB stars reach their position along the branch with different effective temperatures. The total mass deficit of the resulting stars with respect to the RGB progenitors represents the sought-after mass loss.
 Based on this idea, \citet{gratton2010} estimated the RGB mass loss of 98 Galactic GCs, but the presence of multiple stellar populations with different chemical composition provides a significant challenge to their conclusions. 

Indeed, in addition to mass loss, the color and the magnitude of a star along the  HB depend on its age, metallicity, and helium content.
While the majority of GCs hosts stars with the same age and metallicity, the majority of them are composed of two or more stellar populations with different helium content \citep[e.g.][]{dantona_2002, dantona_2005, milone_2018, lagioia_2018}. 
 As a consequence, different parameters with degenerate effects determine the effective temperature of an HB star. In particular, increased helium and RGB mass loss would both increase the star temperature. 

Recent work has introduced an innovative approach to infer the mass loss of the distinct stellar populations in GCs \citep[][]{tailo_2019b,tailo_2019a}. 
These papers are based on the theoretical and empirical evidence that stellar populations with different helium content populate distinct HB regions \citep[e.g.][]{dantona_2002, marino_2011, marino_2014, gratton_2011, dondoglio_2020}.  Once stellar populations are identified along the HB and their helium content is independently constrained from the MS and RGB \citep{milone_2018},  it is possible to disentangle the effect of helium and mass loss along the HB.
 
\citet[][hereafter T20]{tailo_2020} has extended this method to a large sample of 46 GCs. They identified their stellar population with pristine helium abundance (hereafter first generation or 1G)  along the RGB and the HB and inferred the RGB mass loss by using appropriate theoretical models. Similarly, they estimated the mass loss of stars with extreme helium content (hereafter extreme second generation or 2Ge). Tailo and collaborators found that the mass loss of 1G stars changes from one cluster to another and is tightly correlated with the cluster metallicity. Based on these results they defined an empirical mass-loss law for 1G stars.   

In this work, we analyze seven clusters with no evidence of multiple populations, namely NGC\,6426, Palomar\,12, Palomar\,15, Pyxis, Ruprecht\,106, Terzan 7 and Terzan 8. Hence, they are either simple stellar populations or host stars with very small internal helium variations. 
We compare the RGB mass losses inferred from these clusters and from 1G and 2Ge stars of the multiple-population clusters studied by T20. The main goal is to shed light on whether the mass loss law by T20 describes 1G stars of multiple-population GC alone or is a universal property of stellar evolution.

The paper is organized as follows: in \S\,\ref{sec:data_models} we present the photometric catalogues and the stellar evolution models. In \S\,\ref{sec:res} we describe the method to infer mass loss and provide the results for all clusters, in \S\,\ref{sec:res_disc}. Finally, we compare the results from this paper and from T20 and summarize the main findings of the paper in \S\,\ref{sec:Conc}.

\section{Data, data reduction and simulated photometry}
\label{sec:data_models}
The clusters studied in this paper are seven Galactic GCs older than $\sim$9 Gyr, whose HB stars exhibit short color extensions as expected from simple populations.
The sample includes Ruprecht\,106, which is considered the prototype of simple-population clusters as neither high-resolution spectroscopy, nor multi-band photometry show evidence of internal variations in chemical composition \citep[][]{villanova_2012, dotter_2018, lagioia_2019}. 
Pyxis and Palomar 12 are candidate simple-population clusters because their HB span a short color range \citep[][]{milone_2014}. 
 Terzan\,7 is considered a simple-population GC based on the spectroscopic analysis by \citet[][]{sbordone_2005}. In the case of Terzan\,8 high-resolution spectroscopy has revealed that 19 out of 20 analyzed stars  are consistent with 1G stars \citep[][]{carretta_2014}. 
These five clusters have all initial masses \citep[from][]{baumgardt_2018,baumgardt_2019} smaller that $\sim 2 \times 10^{5} M_{\odot}$, which is considered the mass threshold to form multiple populations in GCs \citep[][]{milone2020a}.
In this work we consider these clusters as SSP GC candidates.
We included in the sample Palomar\,15 and NGC\,6426, whose HBs have short color extension but are more massive than $\sim 2 \times 10^{5} M_{\odot}$.
Despite there is no evidence that these two clusters have homogeneous chemical composition, the small color extensions of their HBs suggest that their 2G stars, if present, would not exhibit extreme chemical compositions. 

In addition, we analyzed the CMDs of the candidate simple-population GCs AM\,4, E\,3, Palomar\,1, Palomar\,13 \citep{monaco_2018, milone2020a}. By using literature photometric catalogues \citep[][]{sarajedini_2007, anderson_2008, dotter_2011, milone_2016}, we verified that no HB stars are present in these last four clusters. Nevertheless, we used their photometry to derive other quantities that are relevant for our analysis, including cluster age and the stellar mass at the tip of the RGB. 

 In the following, we summarize the photometric data-set and the stellar models that we employ to derive the RGB mass loss in the seven clusters with HB stars.

\subsection{Photometric dataset}
\label{sec:data_models_data}

To infer the RGB mass loss we derived stellar photometry and proper motions by using two-epoch images collected through the F606W and F814W filters of the Wide Field Channel of the Advanced Camera for Surveys (WFC/ACS) on board {\it HST}.
The main properties of the images are provided in Table \ref{tab:data}.

 Stellar magnitudes and positions have been derived for each exposure separately by using the  img2xym$\_$WFC computer program from \citet{anderson2006a}. 
 In a nutshell, we identified as a candidate star every point-like source whose central pixel has more than 50 counts within its 3$\times$3 pixels and with no brighter pixels within a radius of 0.2 arcsec. The fluxes and positions of all candidate stars have been measured in each exposure by fitting the appropriate effective-PSF model. The stellar magnitudes of all exposures collected through the same filter are then averaged together to provide the best estimates of the instrumental F606W and F814W magnitudes. Instrumental magnitudes are then calibrated to the Vega system by using the photometric zero points provided by the Space Telescope Science Institute website\footnote{https://www.stsci.edu/hst/instrumentation/acs/data-analysis/zeropoints}.

   Stellar position have been corrected for geometrical distortion by using the solution by \citet{anderson2006a} and transformed into a common reference frame based on Gaia early data release 3 \citep[Gaia EDR3,][]{gaia2020a}.
   Stellar coordinates derived from images collected at different epochs are then averaged together and these average positions have been compared with each other to derive the stellar proper motions relative to the average cluster motion \citep[see][for details]{anderson2003a, piotto2012a}. 
   
   These relative proper motions have been transformed into an absolute reference frame by adding to the relative proper motion of each star the average motion of cluster members.
   The absolute GC proper motions are listed in Table \ref{tab:param_dm} and are derived from stars where both Gaia DR3 absolute proper motions and {\it HST} relative proper motions are available.  
   Finally, photometry has been corrected for differential reddening by following the procedure by \citet[][]{milone_2012c} \footnote{The photometric and astrometric catalogs will be available at the   http://progetti.dfa.unipd.it/GALFOR web page and at the  CDS (cdsarc.u-strasbg.fr). }.
   
\begin{figure*}
    \centering
    \includegraphics[height=11.5cm,trim={1cm 5.5cm 12.25cm 4.5cm},clip]{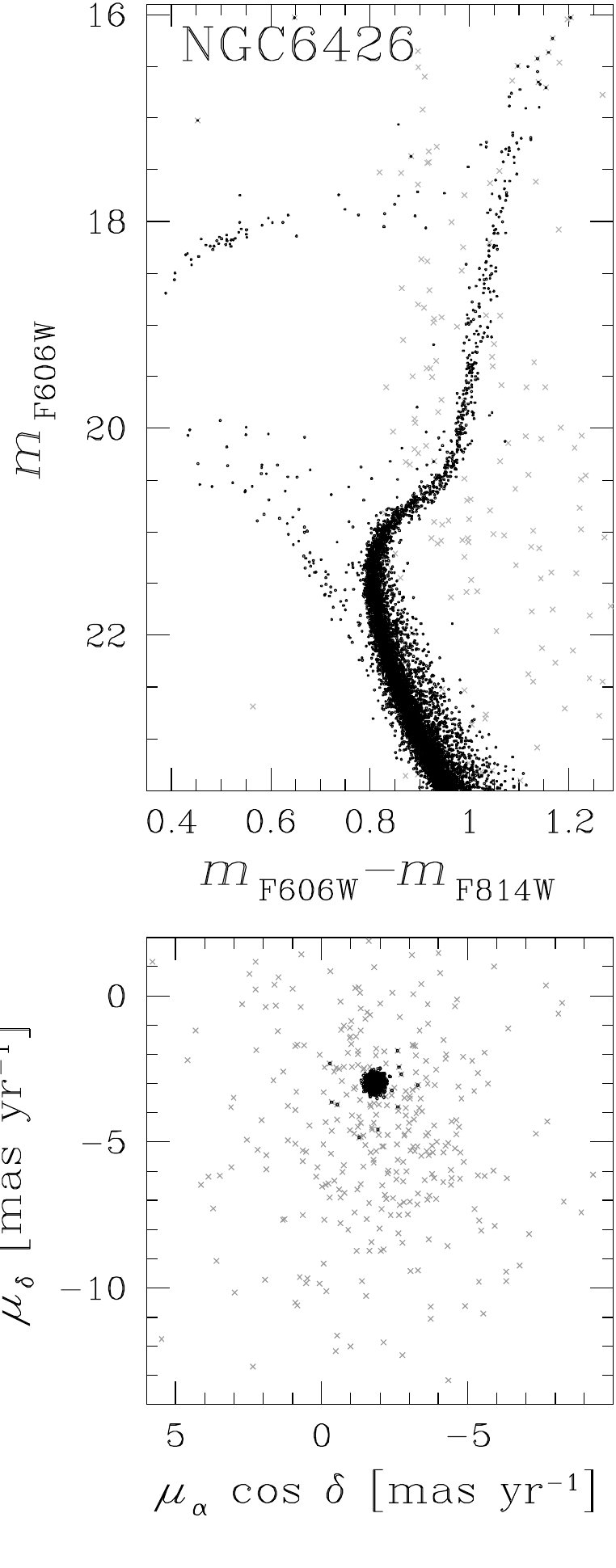}
    \includegraphics[height=11.5cm,trim={2cm 5.5cm 12.25cm 4.5cm},clip]{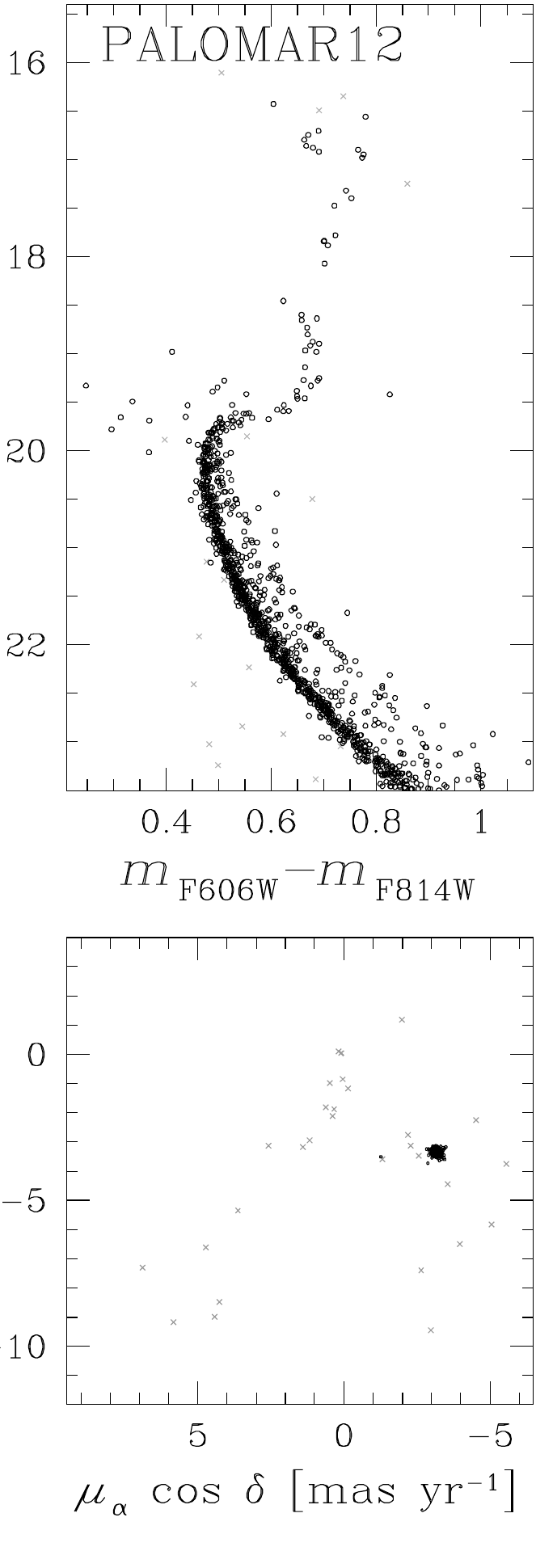}
    \includegraphics[height=11.5cm,trim={2cm 5.5cm 12.25cm 4.5cm},clip]{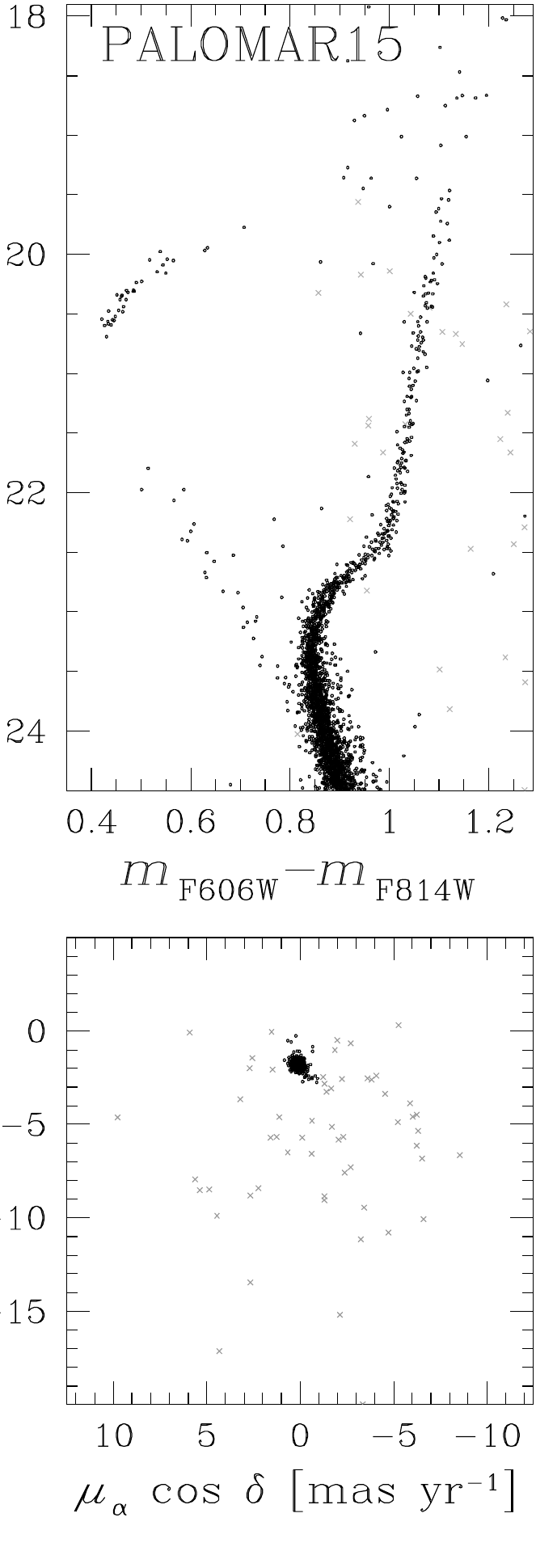}
    \includegraphics[height=11.5cm,trim={2cm 5.5cm 12.25cm 4.5cm},clip]{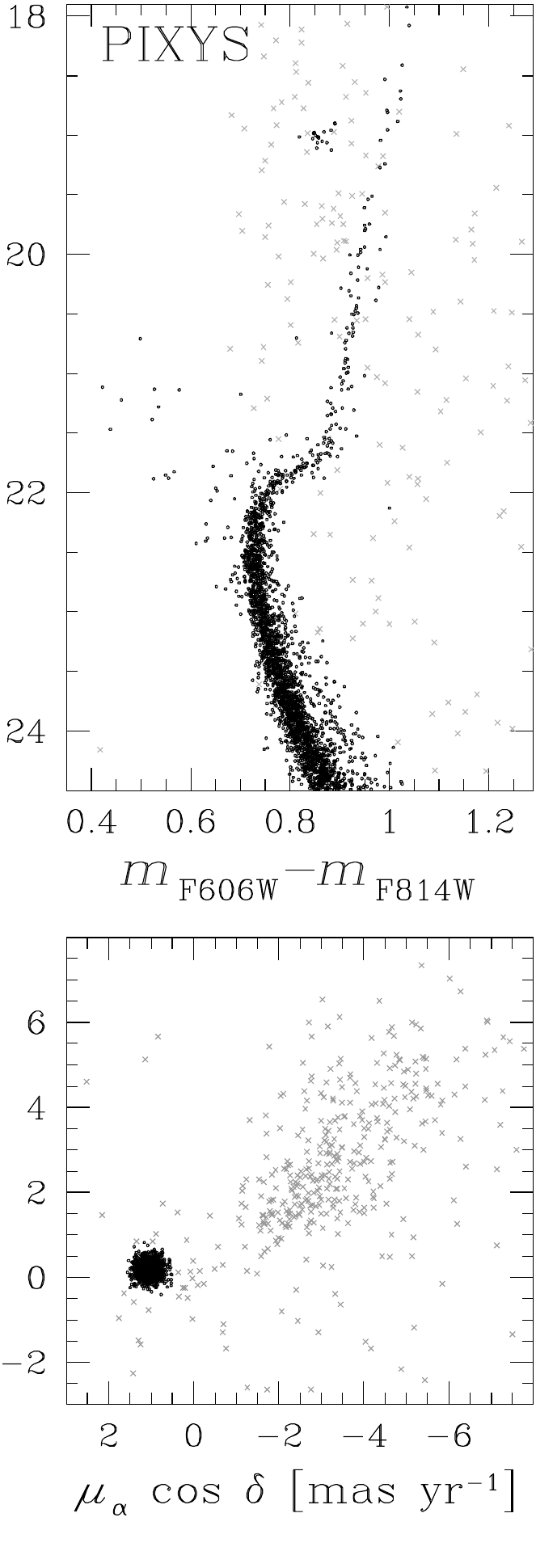}
    \caption{$m_{\rm F606W}$ vs.\,$m_{\rm F606W}-m_{\rm F814W}$ CMDs (upper panels) and vector-point diagrams of proper motions (lower panels) of NGC\,6426, Palomar\,12, Palomar\,15 and Pyxis. Candidate cluster members and field stars are colored black and gray, respectively.}
    \label{pic:CMDs1}
\end{figure*}

\begin{figure*}
    \centering
    \includegraphics[height=11.5cm,trim={.5cm 5.5cm 12.25cm 4.5cm},clip]{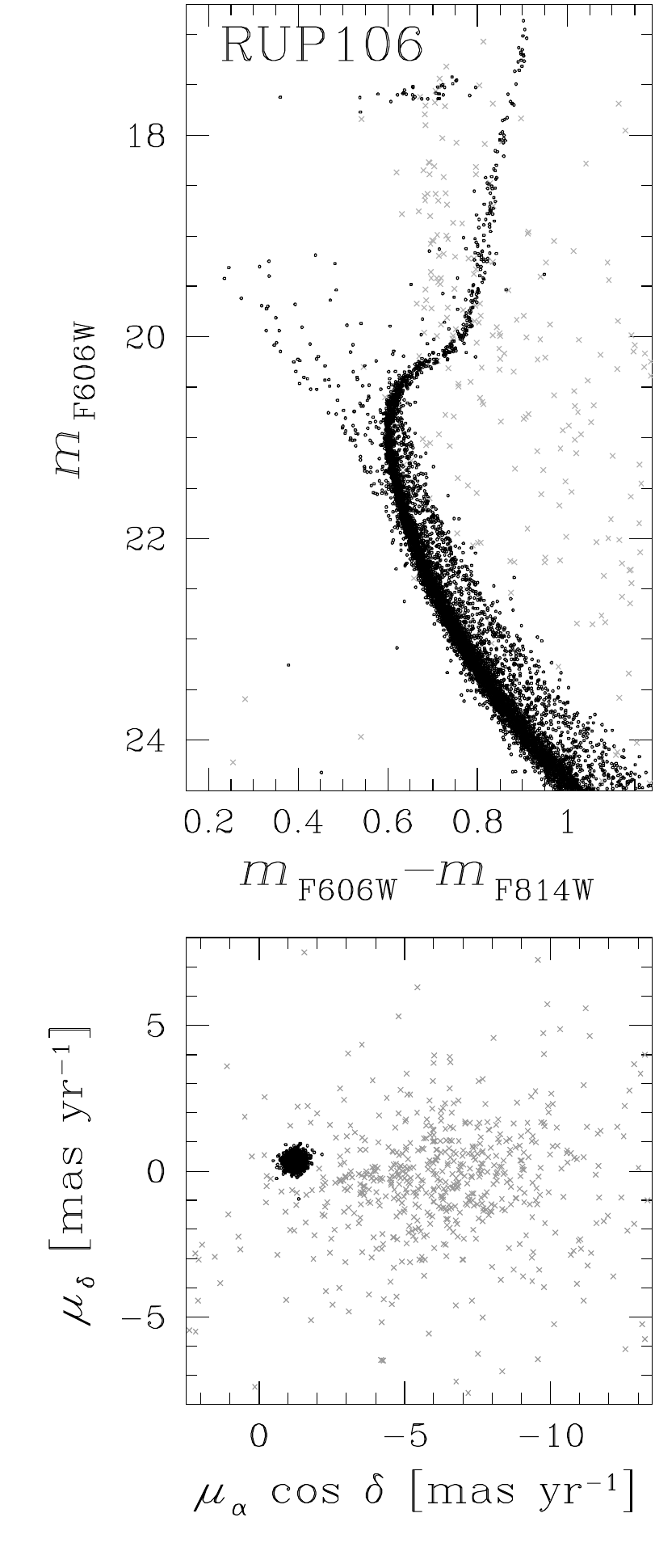}
    \includegraphics[height=11.5cm,trim={2cm 5.5cm 12.25cm 4.5cm},clip]{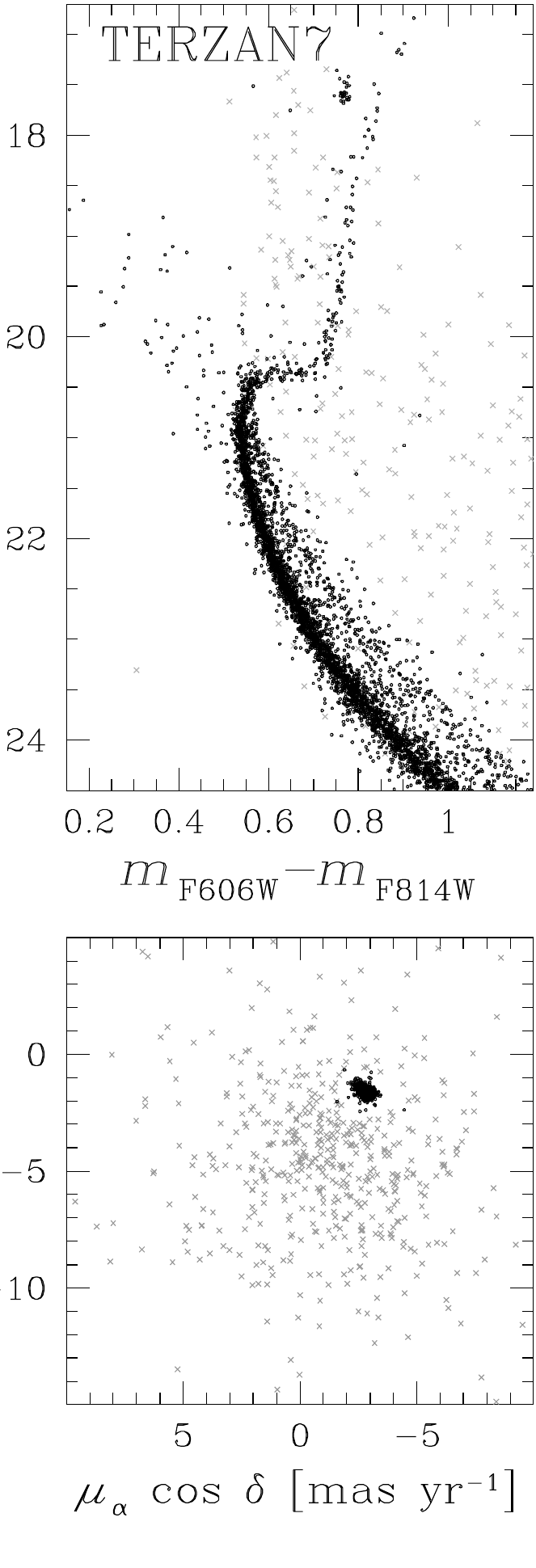}
    \includegraphics[height=11.5cm,trim={2cm 5.5cm 12.25cm 4.5cm},clip]{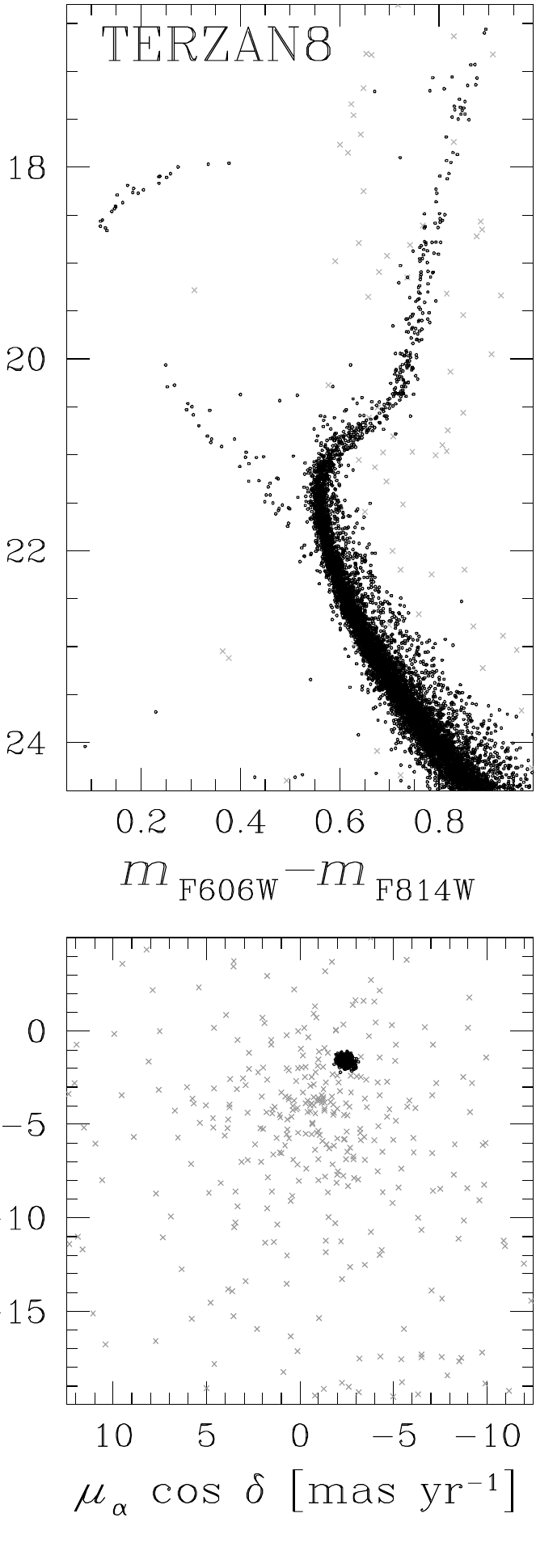}
    \caption{As in Figure \ref{pic:CMDs1} but for Ruprecht\,106, Terzan\,7 and Terzan\,8.}
    \label{pic:CMDs2}
\end{figure*}

   The vector-point diagrams of proper motions and the $m_{\rm F606W}$ vs.\,$m_{\rm F606W}-m_{\rm F814W}$ CMDs corrected for differential reddening of the seven GCs in our sample are plotted in Figures \ref{pic:CMDs1} and \ref{pic:CMDs2}, where we indicated cluster members, selected on the basis of their proper motions \citep[e.g.][]{cordoni2020a}, with black points and field stars with gray crosses.

\subsection{Stellar models}
\label{sec:data_models_models}
We exploited the stellar-evolution models and the isochrones used by T20, which have been computed with the stellar-evolution program ATON 2.0 \citep[][]{ventura_1998,mazzitelli_1999}.  
The grid of models used in this paper includes different ages, metallicity (Z), and helium mass fractions (Y). Specifically, the iron abundance ranges from [Fe/H]=$-$2.44 to $-$0.45  and the adopted values for [$\rm \alpha$/Fe] are 0.0, $+$0.2 and $+$0.4.
In particular, the models with [$\rm \alpha$/Fe]= +0.0 have been calculated specifically for this work. The HB models include a small correction to their helium mass fraction to account for the first dredge up effects. The HB evolution is followed until the end of the helium burning phase. Gravitational settling of helium and metals is not included.

To derive the RGB mass loss of each cluster we compare the CMD of the observed HB stars with a grid of synthetic CMDs, obtained following the recipes of \citet[][and references therein]{dantona_2005}. 
 Briefly, the mass of the each HB star ($\rm M^{HB}$) in each simulation is obtained as: $\rm M^{HB}=M^{Tip}(Z, Y, A) - \Delta M(\mu,\delta)$. Here $\rm M^{Tip}$ is the stellar mass at the RGB tip, which depends on age (A), metallicity (Z) and helium content (Y); $\rm \Delta M$ is the mass lost by the star and described by a Gaussian profile with central value $\mu$ and standard deviation $\delta$. 
 
 Once the value of $\rm M^{HB}$ is obtained the star is placed on its HB track via a series of random extraction and interpolation procedures. Each simulation in the grid is composed of few thousands star to avoid problems due to high variance.
 The values of $\rm M^{Tip}$ are obtained from the isochrones that provide the best fit with the observed CMD. We refer to \cite{dantona_2005,tailo_2016b} and T20 for additional details on the procedure.

%%%%%%%%%%%%%%%%%%%%%%%%%%%%%%%%%%%%%%%%%%%%%%%%%%%%%%%%%%%%%%%%%%%%%%%%%%%     
\begin{table*}
  \caption{Description of the {\it HST} images used in the paper.}

\begin{tabular}{c c c c l l}
\hline \hline
 ID & FILTER  & DATE & N$\times$EXPTIME & PROGRAM & PI \\
\hline
 NGC\,6426 & F606W &  Aug 04 2009 & 45s$+$4$\times$500s       & 11586 & A.\,Dotter \\
 NGC\,6426 & F814W &  Aug 04 2009 & 50s$+$4$\times$540s       & 11586 & A.\,Dotter \\
 NGC\,6426 & F606W &  Aug 09 2016 & 50s$+$3$\times$780s$+$3$\times$781s  & 14235 & S.\,Sohn\\

 Palomar\,12 & F606W &  May 21 2006 & 60s$+$5$\times$340s       & 10775 & A.\,Sarajedini \\
 Palomar\,12 & F814W &  May 21 2006 & 50s$+$5$\times$340s       & 10775 & A.\,Sarajedini \\
 Palomar\,12 & F606W &  Jun 11 2016 & 65s$+$2$\times$544s$+$2$\times$545s$+$548s$+$3$\times$549s  & 14235 & S.\,Sohn\\

 Palomar\,15 & F606W &  Oct 16 2009 & 10s$+$65s$+$8$\times$550s & 11586 & A.\,Dotter \\
 Palomar\,15 & F814W &  Oct 16 2009 & 10s$+$25s$+$55s$+$4$\times$500s$+$4$\times$525s$+$4$\times$560s & 11586 & A.\,Dotter \\
 Palomar\,15 & F606W & Oct 09 2015 & 555s  & 14235 & S.\,Sohn\\
 Palomar\,15 & F814W & Oct 09 2015 & 545s$+$3$\times$546s$+$3$\times$555s  & 14235 & S.\,Sohn\\

 Pixys & F606W &  Oct 11 2009 & 50s$+$4$\times$517s  & 11586 & A.\,Dotter \\
 Pixys & F814W &  Oct 11 2009 & 55s$+$4$\times$547s  & 11586 & A.\,Dotter \\
 Pixys & F606W &  Oct 09 2015 & 50s$+$3$\times$799s$+$3$\times$800s  & 14235 & S.\,Sohn\\

 Ruprecht\,106 & F606W &  Jul 04 2010 & 55s$+$4$\times$550s  & 11586 & A.\,Dotter \\
 Ruprecht\,106 & F814W &  Jul 10 2010 & 60s$+$3$\times$585s$+$586s  & 11586 & A.\,Dotter \\
 Ruprecht\,106 & F606W &  Jul 12 2016 & 60s$+$841s$+$2$\times$842s$+$845$+$2$\times$845s  & 14235 & S.\,Sohn\\

 Terzan\,7 & F606W &  Jun 03 2006 & 40s$+$5$\times$345s       & 10775 & A.\,Sarajedini \\
 Terzan\,7 & F814W &  Jun 03 2006 & 40s$+$5$\times$345s       & 10775 & A.\,Sarajedini \\
 Terzan\,7 & F606W &  May 04 2016 & 45s$+$2$\times$554s$+$2$\times$555s$+$4$\times$556s  & 14235 & S.\,Sohn\\

 Terzan\,8 & F606W &  Jun 03 2006 & 40s$+$5$\times$345s       & 10775 & A.\,Sarajedini \\
 Terzan\,8 & F814W &  Jun 03 2006 & 40s$+$5$\times$345s       & 10775 & A.\,Sarajedini \\
 Terzan\,8 & F606W &  Apr 28 2016 & 45s$+$2$\times$554s$+$2$\times$555s$+$4$\times$556s  & 14235 & S.\,Sohn\\

     \hline\hline
\end{tabular}
  \label{tab:data}
 \end{table*}
%% %%%%%%%%%%%%%%%%%%%%%%%%%%%%%%%%%%%%%%%%%%%%%%%%%%%%%%%%%%%%%%%%%%%%%%%%%%

\section{Constraining the RGB mass loss}
\label{sec:res}
 In this section we summarize the procedure to derive the RGB mass loss in simple-population clusters, using Palomar 15 as a template. After extending the same analysis to the other analyzed GCs, we will discuss the results.
 
\subsection{Example case: Palomar 15}
\label{sec:res_6426}

\begin{figure*}
    \centering
    \includegraphics[width=2\columnwidth]{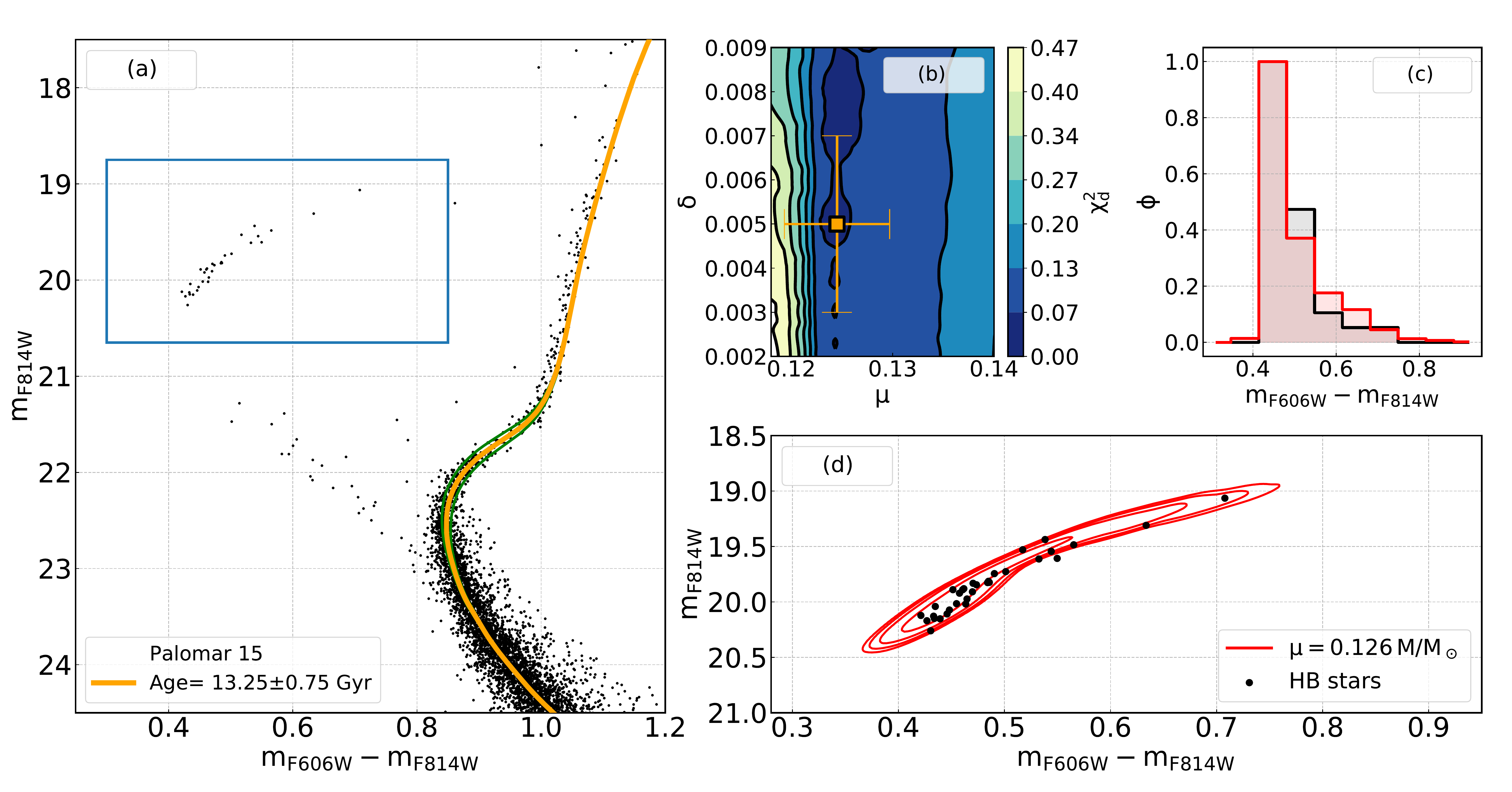}
    \caption{
    Summary of the procedure we follow to obtain the RGB mass loss for Palomar\,15. In the panel (a) we compare the observed CMD of cluster member with the best-fit isochrone (yellow line) and the isochrones with ages of $\pm$1$\sigma$ Gyr from the best age estimate (green line). The blue rectangle highlights HB stars.
    Panel (b) shows the $\rm \chi^2_d$ density plot in the mass-loss dispersion vs.\, mass loss plane. The best fit values are marked by the orange square. The histograms of the color distribution of observed (black) and simulated HB stars are shown in panel (c), while panel (d) shows the observed HB and contours corresponding to the best fit simulated HB. See text for details.   
    }
    \label{pic:cmd_hist}
\end{figure*}

To infer the RGB mass loss experienced by HB stars we follow the procedure introduced by T20 \citep[see also][]{tailo_2019b, tailo_2019a} and illustrated in Figure \ref{pic:cmd_hist} for Palomar\,15. This procedure is based on the comparison between the observed HB stars and a grids of simulated HBs.

 At odds with T20, who studied clusters with multiple populations, our sample consists in candidate simple-population GCs. Hence, we assumed that all their stars have the same chemical composition and adopted pristine helium mass fraction for all clusters.

The first step is to evaluate the age of the population, needed as input to generate the synthetic HB grids. We do that via the isochrone fitting of the turn off region. We produce an array of isochrones with [Fe/H]=$-$2.07, [$\rm \alpha$/Fe]=+0.4 \citep[following the indication by][]{kirby_2008} and Y=0.25, with age ranging from 8.0 to 14.0 Gyr in steps of 0.25 Gyr. We adopt E(B$-$V)=0.40 and (m$-$M)$_{\rm V}$=19.51 from the 2010 version of the \citet[][]{harris_1996} catalogue\footnote{  https://heasarc.gsfc.nasa.gov/w3browse/all/globclust.html or  https://www.physics.mcmaster.ca/$\sim$harris/mwgc.dat}. 

The isochrone that provides the best match with the turn-off region in the $\rm m_{F814W}$ vs $\rm m_{F606W}-m_{F814W}$ CMD (orange isochrone in Figure \ref{pic:cmd_hist}a) gives us the best estimate for cluster age, 
in this case  $13.25\pm0.75$ Gyr. The uncertainty corresponds to %
 the age range that allows
 the isochrones to envelope 68.27\% of stars in the turn off region. 
 
We verified that our age estimate is not significantly affected by unresolved binaries and blue stragglers (BSS). To investigate the possible effect of binaries and BSSs on the inferred cluster ages, we simulated two mock CMDs of Palomar 15 with the same age and metallicity values adopted here, but different fraction of binaries and BSSs. In the first CMD we assumed no binaries and BSSs, while in the second one the same fraction of binaries derived by \citet{milone_2016} and the same number of BSSs as observed in the actual CMD. We derive the GC age of both simulated CMDs by using the same methods described above and we obtain that the age values are consistent within 0.25 Gyr. Hence, we conclude that binaries and BSSs do not affecr our age determination.

Our second step is to identify, by eye, the HB stars in the cluster.  The selected HB stars of Palomar\,15 are enclosed in the blue rectangle of Figure \ref{pic:cmd_hist}a. We take extra care to verify that the stars are identified on the HB in both $\rm m_{F814W}$ vs $\rm m_{F606W}-m_{F814W}$ and $\rm m_{F606W}$ vs $\rm m_{F606W}-m_{F814W}$ CMDs. 

The selected HB stars are compared with appropriate grids of simulated HBs corresponding to different values of mass loss ($\mu$) and mass-loss dispersion ($\delta$; see T20 for details). 
In the case of Palomar\,15 $\rm \mu$ varies from 0.010 $\rm M_\odot$ to 0.180 $\rm M_\odot$ in steps of 0.003 $\rm M_\odot$, and $\delta$ ranges from 0.002 $\rm M_\odot$ to 0.009 $\rm M_\odot$ in steps of 0.001 $\rm M_\odot$.

For each simulation in the grid we compare the normalized histogram of the color distribution of observed and simulated stars. 
To quantify the goodness of the fit, we calculate the $\chi$-squared distance between the two histograms, $\rm \chi^2_d$ \citep[see][and T20]{chisq_cite}. 
 The resulting density map of $\rm \chi^2_d$ values in the $\rm \mu$ vs.\,$\delta$ plane is plotted in Figure \ref{pic:cmd_hist}b. The best fit simulation is then the one that minimizes the $\rm \chi^2_d$ and is indicated with the orange square on the map of Figure \ref{pic:cmd_hist}b.

We evaluate the uncertainty on our estimates of mass loss and mass-loss dispersion 
 by means of bootstrapping. We generated 5,000 realizations of the HB in Palomar\,15 and performed the comparison with the grid of simulated HBs on each iteration.  To estimate the uncertainties on mass loss and mass loss-dispersion we first considered the standard deviation of the results.
 
Moreover, we added the contribution to the error from the uncertainties on cluster age, metallicity, and reddening. To do this, we derived mass loss by using the same procedure above but by changing cluster age by 0.75 Gyr, iron abundance by 0.10 dex and reddening by E(B$-$V)=0.015 mag. In clusters without spectroscopic determination of $\alpha$-element abundance, we also accounted for the effect of a variation in [$\alpha$/Fe] by 0.2 dex.
By adding in quadrature all the contributions to the total error, we obtain the final estimate for the mass loss of Palomar\,15: $\rm \mu=0.126\pm 0.030 M_\odot$.

The best-fit isochrone provides the mass at the RGB tip,
$\rm M^{Tip} = 0.783\, M_\odot$ and for the HB stars  ($\rm M^{HB} = 0.674 \pm 0.030\,M_\odot$). 
 The complete list of parameters inferred from the procedure illustrated for Palomar\,15 are listed in  in Table \ref{tab:param_dm} for all studied clusters.

For completeness, we compare the histogram of the color distribution of the best fit simulation, i.e. the one whose values of $\rm \mu$ and $\rm \delta$ minimize $\rm \chi^2_{d}$, with the corresponding histogram from observed HB stars (Figure \ref{pic:cmd_hist}c). Finally, in the panel d of Figure \ref{pic:cmd_hist} we superimposed on the observed CMD the contours of the best-fit simulation. In the panel \ref{pic:cmd_hist}d, the contour lines delimit the regions of the simulated CMD including (starting from the outermost region) 98,95,80,60\% of stars.

\section{Mass loss and HB mass in candidate simple-population GCs}
\label{sec:res_sample}

\begin{figure*}
    \centering
    \includegraphics[width=0.66\columnwidth]{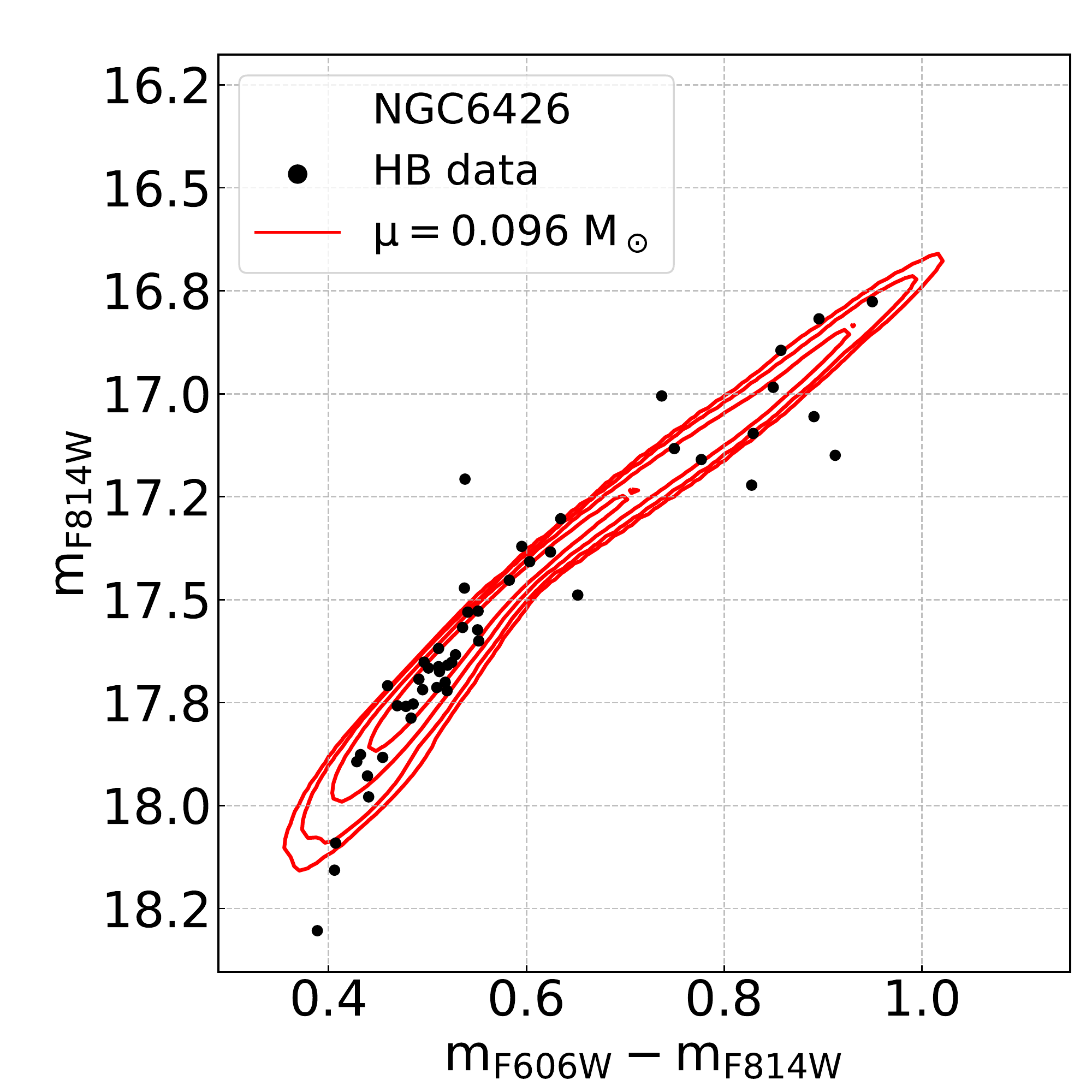}
    \includegraphics[width=0.66\columnwidth]{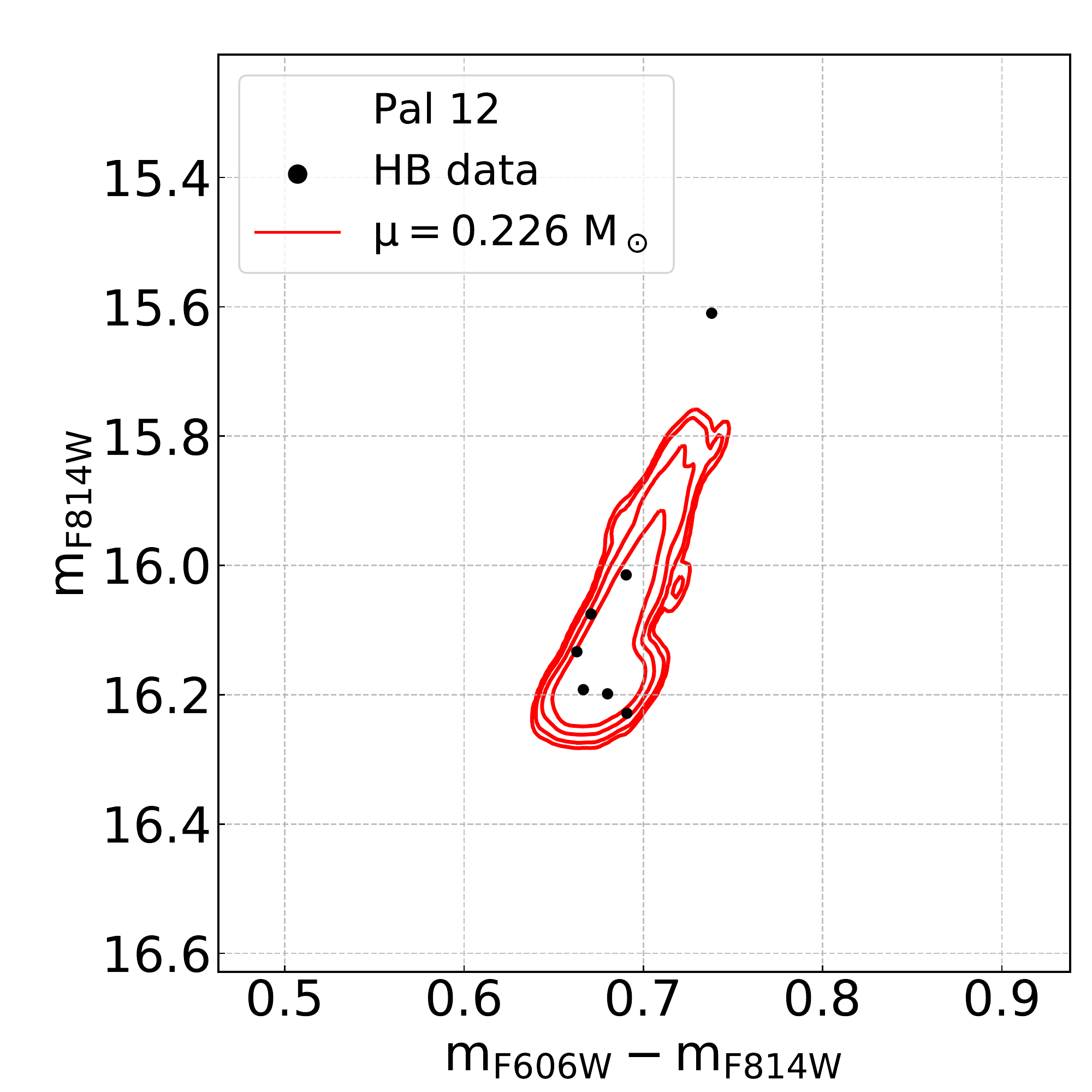}
    \includegraphics[width=0.66\columnwidth]{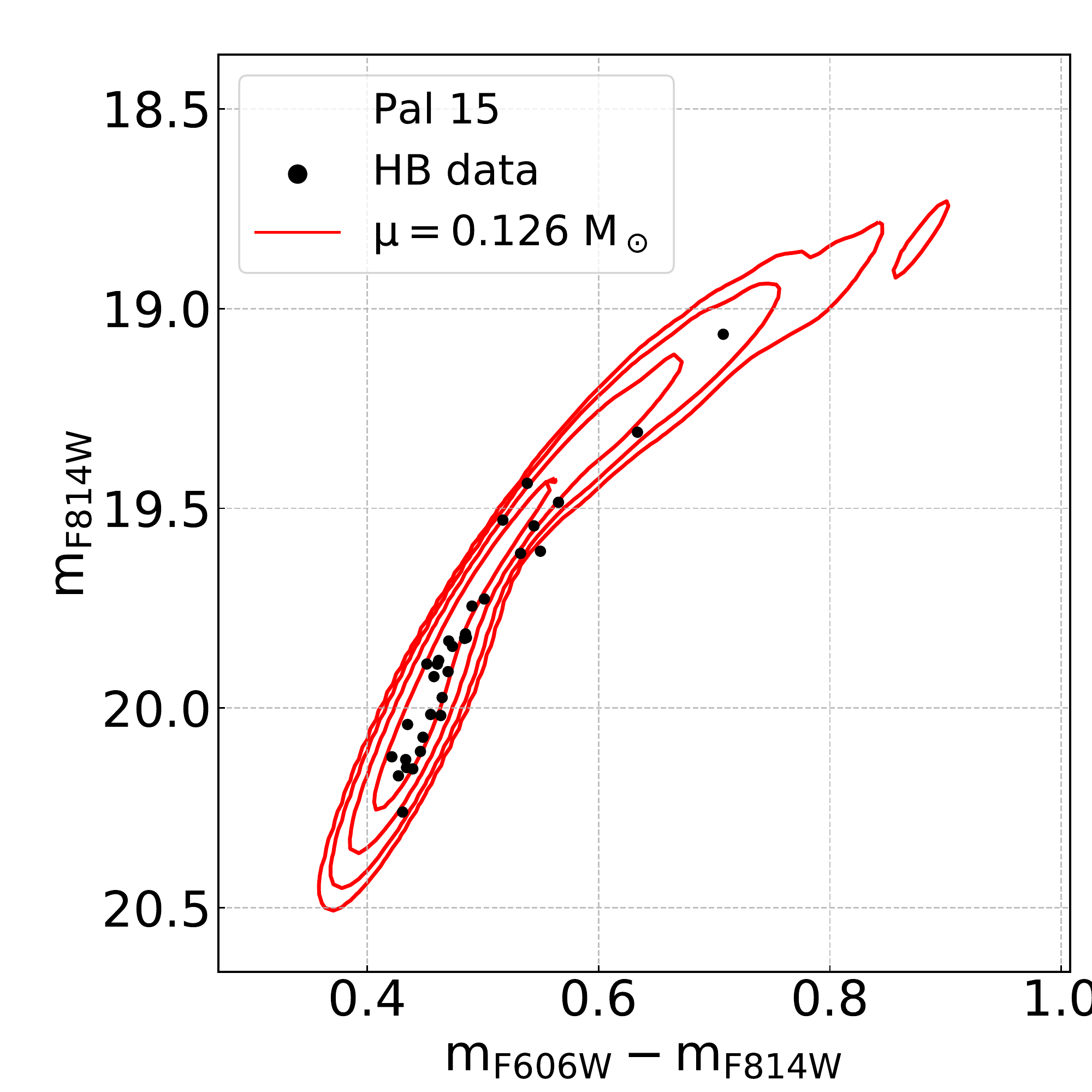}
    \includegraphics[width=0.66\columnwidth]{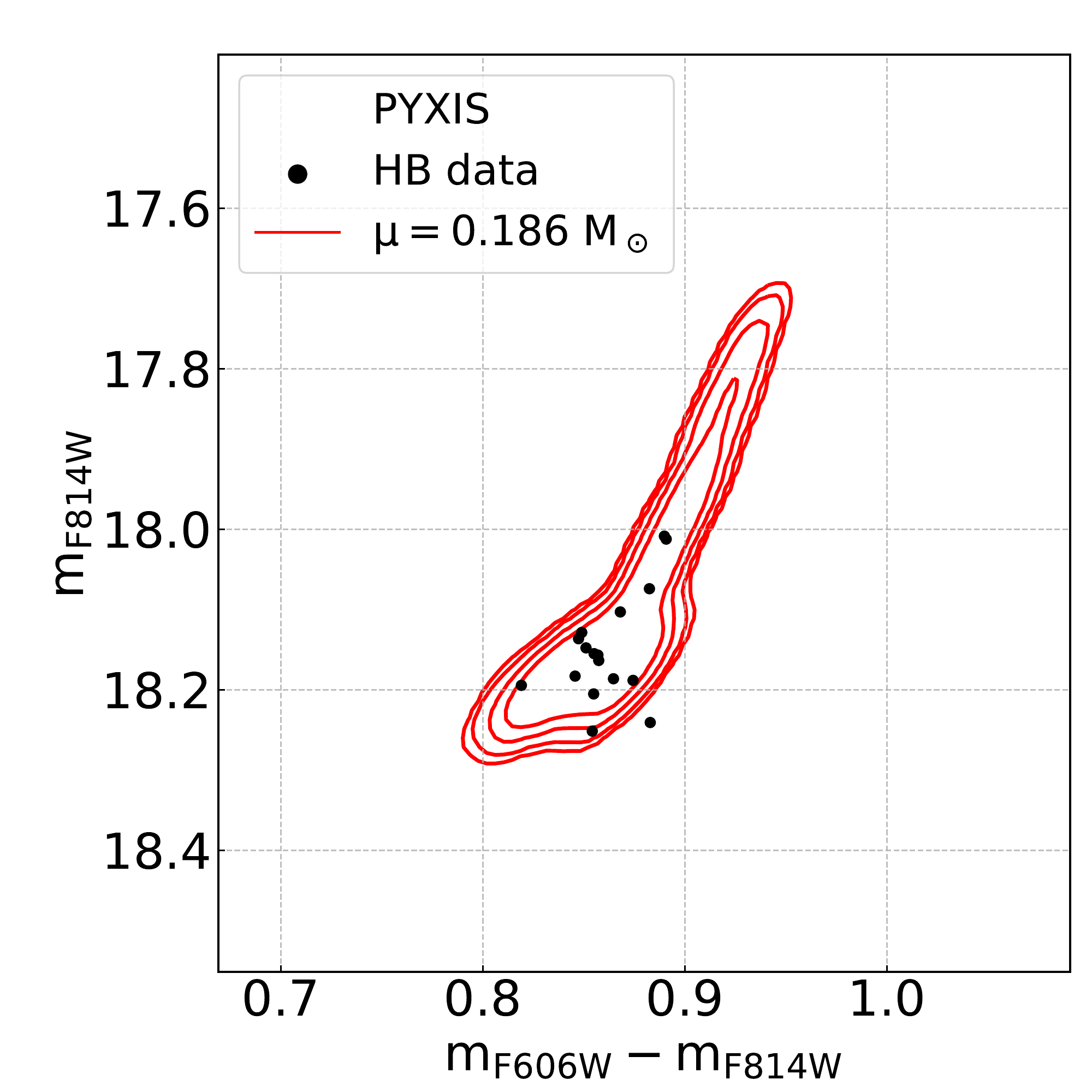}
    \includegraphics[width=0.66\columnwidth]{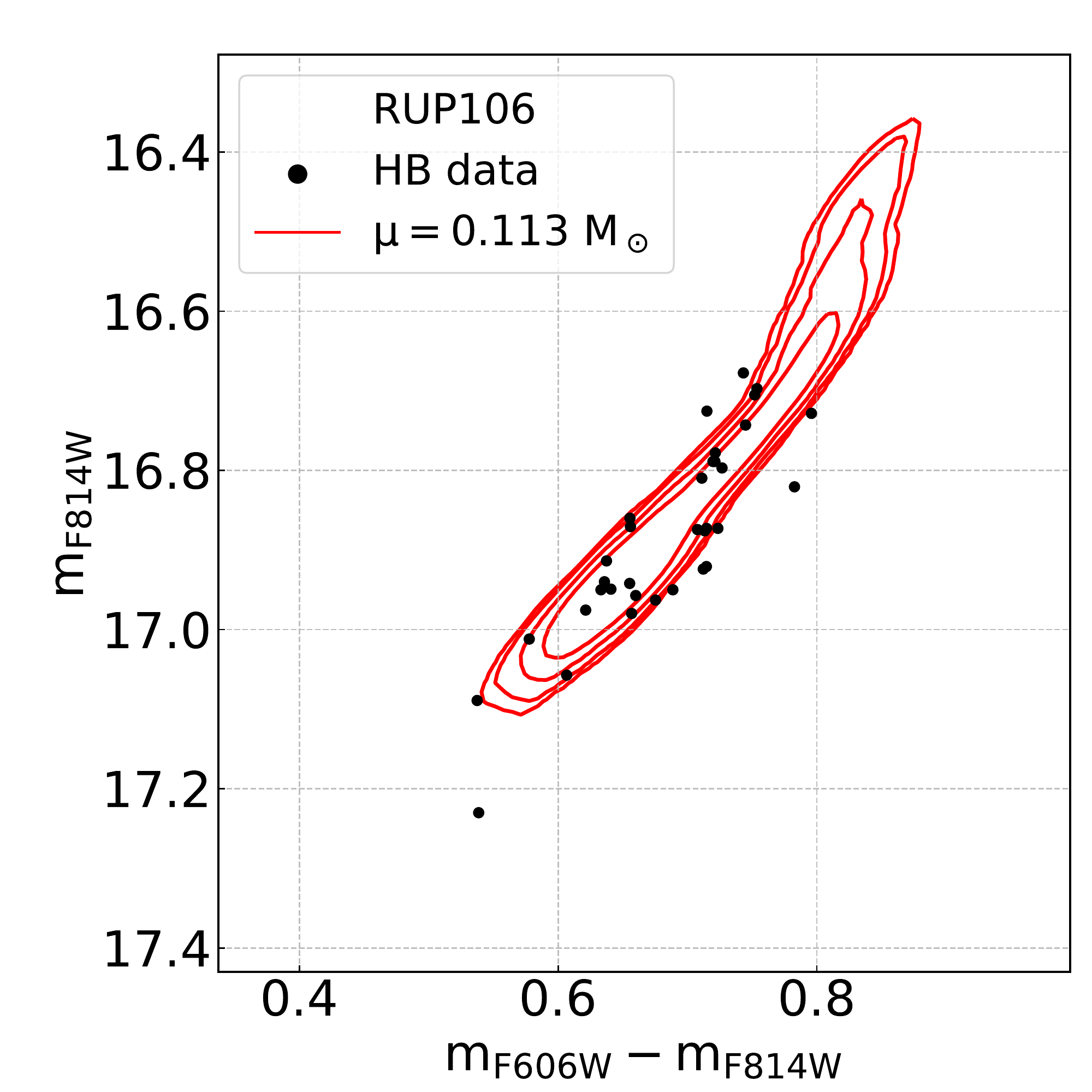}
    \includegraphics[width=0.66\columnwidth]{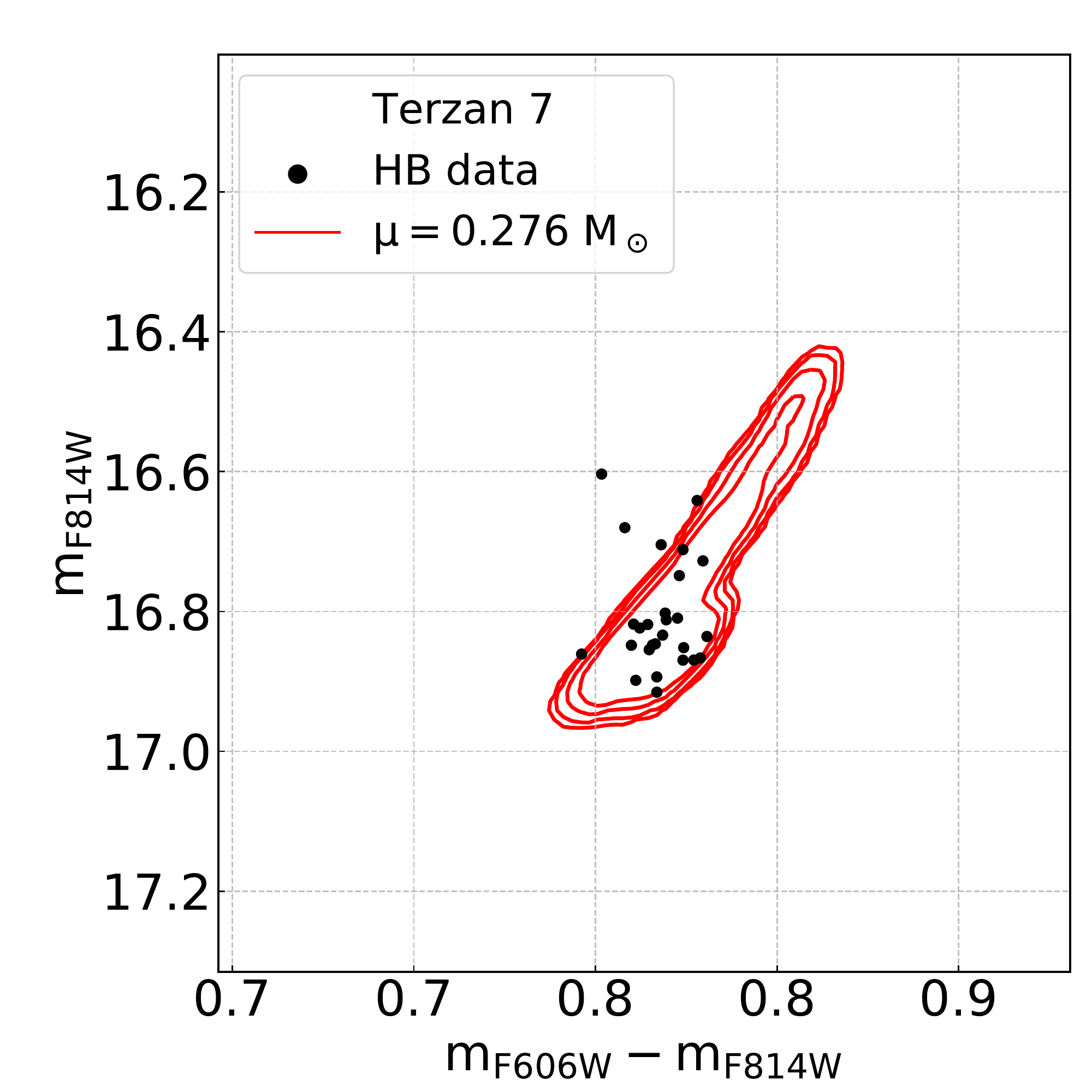}
    \includegraphics[width=0.66\columnwidth]{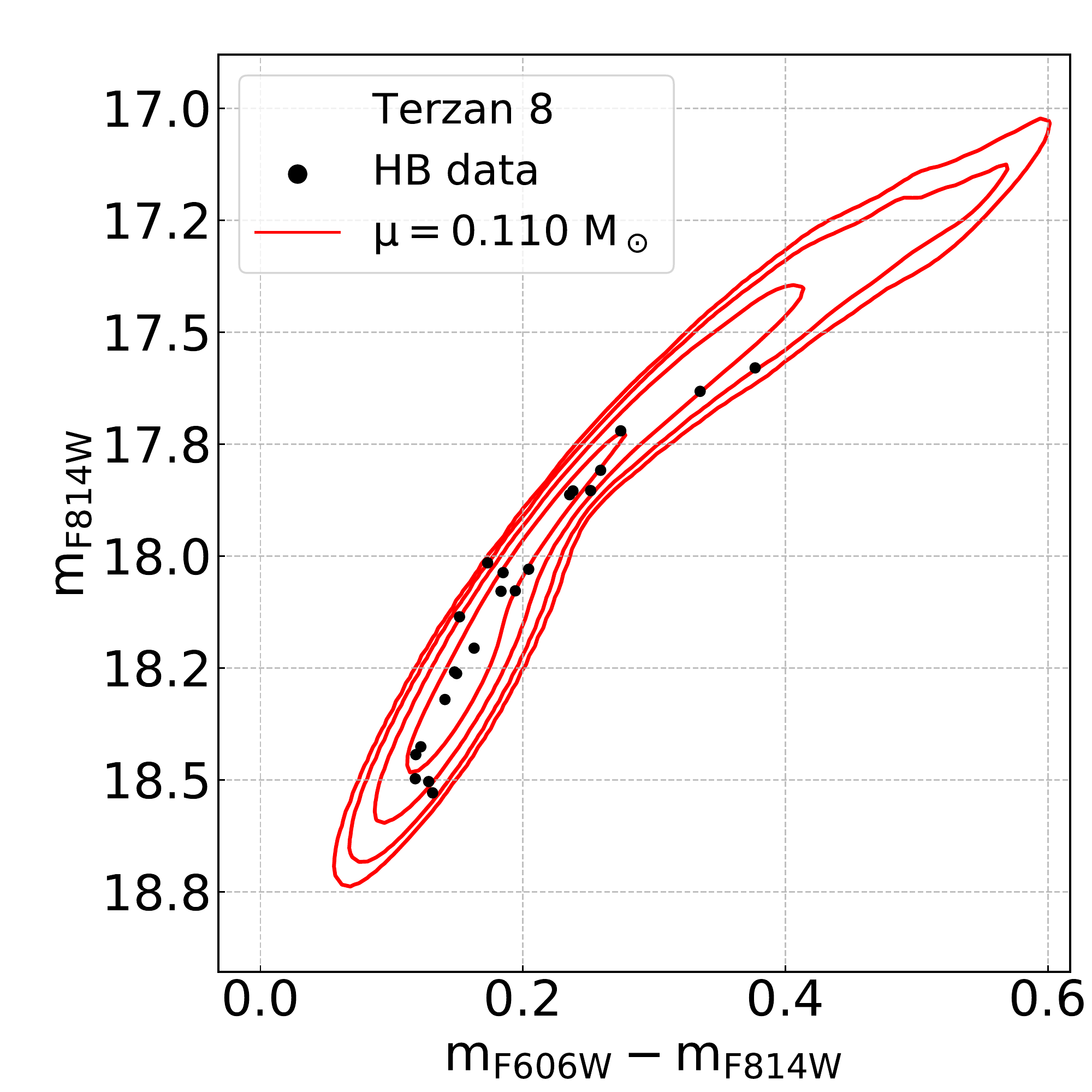}
    \caption{
    The analysed GCs in alphabetical order. We report in each panel the $\rm m_{F814W}$ vs $\rm m_{F606W}-m_{F814W}$ CMD of the HB stars together with the contour plot of the best fit simulation.  The average mass losses of the best fit simulations are quoted in the insets.
    }
    \label{pic:showcase}
\end{figure*}

The procedure from T20, summarized in the previous section, has been extended to the entire sample of eleven Galactic GCs that are candidate to host a simple stellar population. 

The analyzed GCs are listed in Table \ref{tab:param_dm}, together with the values of [Fe/H], [$\rm \alpha$/Fe], E(B$-$V) and (m$-$M)$_{\rm V}$ used for their analysis. Specifically, the values of reddening and distance modulus are from the 2010 version of the \citet[][]{harris_1996} catalog, while the values of [Fe/H] and [$\rm \alpha$/Fe] are taken from various literature sources and are derived from spectroscopy (see Table \ref{tab:param_dm} for the complete list of references).

 Since no spectroscopic determination for $\alpha$ elements are available for AM\,4, Palomar\,15 and Pyxis, we fixed the [$\alpha$/Fe] values based on their metallicity as suggested by \citet[][]{kirby_2008}. Hence, we assumed [$\alpha$/Fe]=0.2 for AM\,4, [$\alpha$/Fe]=0.4 for Palomar\,15 and [$\alpha$/Fe]=0.2 for Pyxis. We also include in the error budget for these clusters the effects of a possible shift of 0.2 in [$\alpha$/Fe] that stems from the results of \citet{kirby_2008}. 
 
  The complete showcase of examined HBs with their best fit simulations is plotted in Figure \ref{pic:showcase}, while the values of $\rm \mu$, $\rm \delta$ and $\rm M^{HB}$ derived from the analysis described in Section 3.1 are listed in Table \ref{tab:param_dm}.

\begin{table*}
    \centering
    \caption{
    Parameters of the GCs analysed in this work. The values of iron abundances, reddening and distance modulus are taken from the 2010 version of the \citet[][]{harris_1996} catalog. The average GC proper motions are derived in this paper based on Gaia EDR3 motions. Cluster age, mass at the RGB tip, RGB mass loss, mass-loss spread and average HB mass are derived in this paper.  Sources for [$\alpha$/Fe]:
    (a) \citet{sbordone_2005},(b) \citet{dias_2015},(c) \citet{monaco_2018},(d) \citet{cohen_2004},(e) \citet{brown_1997},(f) \citet[][and references therein]{dotter_2018},(g)\citet{pritzl_2005},(h) \citet{jahandar_2017}, (i)\citet{Koch_2019}, (l) from the indications in \citet{kirby_2008}.
    }
    \begin{adjustbox}{width=2\columnwidth,center}
    \begin{tabular}{lccccccccccc}
    \hline
    \hline
    ID  & [Fe/H]&[$\alpha$/Fe]&E(B-V)(mag)&(m-M)$_{\rm V}$(mag) &Age (Gyr)&$\rm M^{Tip}/M_\odot$&$\rm \mu/M_\odot$&$\delta/M_\odot$&$\rm \bar{M}^{HB}/M_\odot$ & $\mu_{\alpha} cos{\delta}$ (mas yr$^{-1}$) & $\mu_{\delta}$ (mas yr$^{-1}$)\\
    \hline
    NGC\,6426        & $-$2.15& 0.4$^b$     & 0.36 & 17.68 &$13.50\pm1.00$ & 0.782 &    $0.096\pm0.029$  &  $0.008\pm0.002$&    $0.693\pm0.029$ & $-$1.85$\pm$0.02    & $-$2.98$\pm$0.02 \\
    PALOMAR\,12      & $-$0.85& 0.0$^{d,e}$ & 0.02 & 16.46 &$10.00\pm0.50$ & 0.909 &    $0.226\pm0.035$  &  $0.003\pm0.001$&    $0.683\pm0.035$ & $-$3.14$\pm$0.04    & $-$3.33$\pm$0.03 \\
    PALOMAR\,15      & $-$2.07& 0.4$^l$     & 0.40 & 19.51 &$13.25\pm0.75$ & 0.783 &    $0.126\pm0.030$  &  $0.005\pm0.002$&    $0.657\pm0.030$ & $-$0.60$\pm$0.19    & $-$1.24$\pm$0.20 \\
    PYXIS            & $-$1.20& 0.2$^l$     & 0.21 & 18.63 &$11.00\pm0.75$ & 0.860 &    $0.186\pm0.045$  &  $0.002\pm0.001$&    $0.674\pm0.045$ &    1.04$\pm$0.05    &    0.18$\pm$0.06 \\
    RUPRECHT\,106    & $-$1.68& 0.0$^{e,f}$ & 0.20 & 17.25 &$11.00\pm0.50$ & 0.831 &    $0.113\pm0.035$  &  $0.004\pm0.002$&    $0.718\pm0.035$ & $-$1.24$\pm$0.02    &    0.34$\pm$0.02 \\
    TERZAN7          & $-$0.60& 0.0$^a$     & 0.07 & 17.01 &$ 9.00\pm0.50$ & 0.954 &    $0.276\pm0.045$  &  $0.006\pm0.002$&    $0.678\pm0.045$ & $-$2.90$\pm$0.02    & $-$1.66$\pm$0.02 \\
    TERZAN8          & $-$2.16& 0.4$^b$     & 0.12 & 17.47 &$13.50\pm0.50$ & 0.782 &    $0.110\pm0.022$  &  $0.005\pm0.002$&    $0.672\pm0.022$ & $-$2.42$\pm$0.02    & $-$1.59$\pm$0.02 \\
    \hline
    AM4              & $-$1.30& 0.2$^l$     & 0.05 & 17.69 &$12.25\pm0.75$ & 0.829 &    ---              &    ---          &    ---	        &    ---          &    ---   \\
    E03              & $-$0.83& 0.2$^c$     & 0.30 & 15.47 &$11.50\pm1.00$ & 0.899 &    ---              &    ---          &    ---	        &    ---          &    ---   \\
    PALOMAR\,1       & $-$0.65& 0.0$^h$     & 0.15 & 15.70 &$ 9.50\pm1.25$ & 0.926 &    ---              &    ---          &    ---	        &    ---          &    ---   \\
     PALOMAR\,13     & $-$1.88& 0.4$^i$     & 0.05 & 17.23 &$11.00\pm0.50$ & 0.836 &    ---              &    ---          &    ---            &    ---          &    ---    \\
    \hline
    \hline
    \end{tabular}
    \end{adjustbox}
    \label{tab:param_dm}
\end{table*}

\begin{figure}
    \centering
    \includegraphics[width=
    \columnwidth]{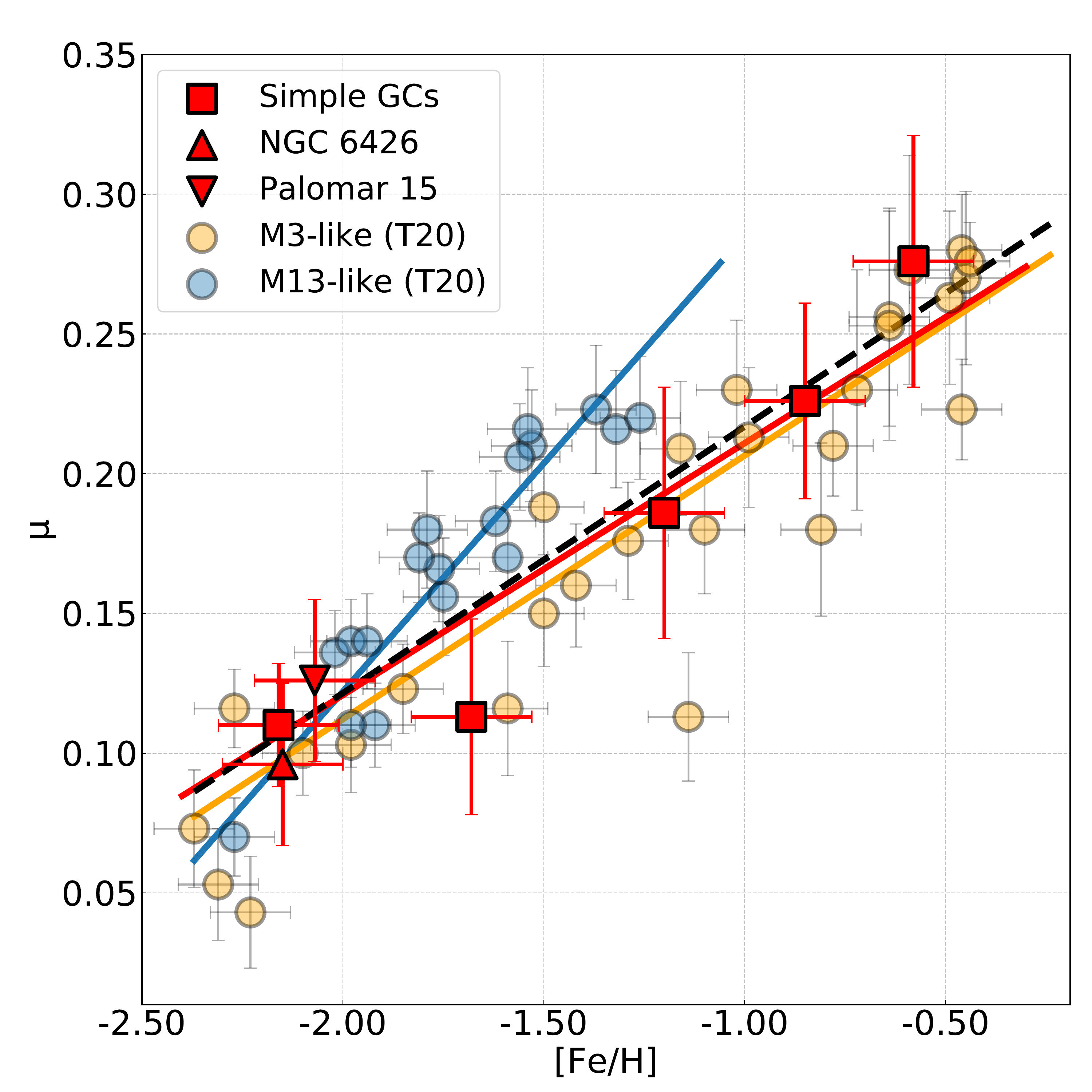}
    \caption{   
     Mass loss ($\rm \mu$) of the HB stars in the simple population GCs as function of their [Fe/H] values. The red line is the best-fit straight line. In the background we plot as orange and blue points the mass loss of the 1G stars for the M3- and M13-like GCs from T20, respectively, and their best fit lines. The black dashed lines is the general relation from T20.
    }
    \label{pic:mu_feh}
\end{figure}

Our results show that mass loss changes from cluster to cluster and ranges from $\rm \mu\sim 0.10\rm M_\odot$ in NGC\,6426 to  $\rm \mu\sim 0.30\rm M_\odot$ in Palomar 1.
We plot in Figure \ref{pic:mu_feh}, as red squares, the values of mass loss for the studied GCs as a function of [Fe/H]. Clearly, mass loss correlates with metallicity as demonstrated by the high values of the Spearman's rank and the Pearson correlation coefficient ($\rm R_s=0.93$ and $\rm R_p=0.95$, respectively). As the identification of NGC\,6426 and Palomar 15 as simple-population cluster is not certain, we will represent them with a different symbol in this and later figures. The points are fitted with a least-squares straight line (red continuous line in Figure \ref{pic:mu_feh}):
 \begin{equation}\label{eq:eq1}
     \rm \mu = (0.090\pm 0.013) \times [Fe/H] + (0.301\pm 0.026) M_\odot
 \end{equation}
 The mean dispersion around the best fit line is 0.019\,$\rm M_\odot$. We report a summary of the parameters in the best-fit relation in Table \ref{tab:fits}.

\section{A universal mass-loss law for Population II stars?}
\label{sec:res_disc}

In their recent  paper, T20 constrained the RGB mass loss of the distinct stellar populations of 46 galactic GCs with multiple populations. In particular, they investigated the mass loss of 1G stars and find a linear relation between the RGB mass loss and the iron abundance of the host GC.
The comparison between the findings of our paper and the results from Tailo and collaborators for 1G stars (Figure \ref{pic:mu_feh})  reveals that simple-population clusters  follow a similar distribution in the mass-loss metallicity plane as 1G stars of multiple-population GCs. 
In particular,  the best-fit line of simple-population GCs described by Equation \ref{eq:eq1} is almost coincident with the corresponding relation discovered by T20 for the entire sample of 1G stars, $\rm \mu = (0.095\pm 0.006) \times [Fe/H] + (0.313\pm 0.011)$ $M_\odot$, black dashed-line of Figure \ref{pic:mtip_feh}. 
In contrast, the RGB mass loss of simple-populations clusters do not match that of 2G stars with extreme helium contents in GCs with similar metallicities.

This evidence suggests that the relation between mass-loss and metallicity by T20 does not depend on the presence of multiple populations in GCs but may be a standard stellar evolutionary property.
 This conclusion is corroborated by the results by \citet{salaris_2013} and \citet{savino_2019} who found a similar mass-loss law in dwarf galaxies. 
 By combining the results by T20 for 1G stars in 46 GCs and those in this paper for simple-population clusters we derive the improved relation: $\rm \mu = (0.095\pm 0.006) \times [Fe/H] + (0.312 \pm 0.011) M_{\odot}$, where the dispersion is $\sim$0.03 $\rm M_\odot$ and the correlation coefficients are $\rm R_s \sim R_p\sim0.89$. As a matter of fact this is equal to the general relation in T20 (black-dashed line in Figure \ref{pic:mu_feh}). We report this general relation in Table \ref{tab:fits}.
 
Recent works, based on asteroseismology, provide mass loss estimates in star clusters.
\cite{Miglio_2016} used Kepler data of seven RGB and one red-HB stars of the GC M\,4 to derive stellar masses. The resulting mass loss, estimated as the mass difference between HB and RGB stars, ranges from  $\sim$0.0 to $\sim$0.2 M$_{\odot}$, depending on the adopted scaling relation. The latter value alone is consistent with the mass-loss inferred by T20 for the 1G stars of M\,4. 

On the other hand, similar studies on the old, metal rich open cluster NGC\,6791 suggest a moderate RGB mass loss for this Galactic open cluster \citep[$\mu=0.09 \pm0.03$ (random) $\pm 0.04$ (systematic) ,][and references therein]{Miglio_2012}, whereas RGB and red clump stars of the $\sim$2.5 Gyr old cluster NGC\,6819 are consistent with sharing the same masses \citep[see also][]{Handberg_2017}.

Both NGC\,6791 and  NGC\,6819 ([Fe/H]$\geq$+0.3 and [Fe/H]$\sim$0.0 respectively) are more metal rich than the GCs studied in this paper, thus preventing us from a proper comparison. However, we note that the mass losses inferred by Miglio and collaborators for NGC\,6791 and  NGC\,6819 are much smaller than those of most metal rich GCs.
This difference may imply that the relation between mass loss and metallicity inferred for GCs can not be extrapolated to Population I stars, being valid up to [Fe/H]$\sim$-0.5. As an alternative, uncertainties in stellar evolution models and/or in asteroseismology scale relations can contribute to the discrepancy between the results based on Kepler data and those of this paper.  

 \subsection{Mass loss as second parameter of the HB morphology}

 T20 identified two groups of GCs with different HB morphology: a group of GCs, that, similarly to M\,3, exhibit the red HB (M\,3-like GCs) and a the group of M\,13-like GCs with the blue-HB alone \citep[see also][]{milone_2014}. M\,3-like and M\,13-like GCs are represented with orange and azure colors, respectively, in Figure \ref{pic:mu_feh}.
 The two groups of M\,3-like and M\,13-like GCs define distinct trends in the $\mu$ vs.\,[Fe/H] plane, with M\,13-like clusters having higher values of $\mu_{\rm 1G}$ than M\,3-like clusters with similar iron content. 
 
Clearly, the best-fit line of simple-population clusters is in agreement within one-$\sigma$ with the corresponding relation of M\,3-like GCs  ($\rm \mu = (0.094\pm 0.007) \times [Fe/H] + (0.302\pm 0.011)$, orange line), but exhibits a different slope than the best-fit line defined by M\,13-like GCs. 

 \begin{table}
    \centering
	\caption{Linear fits in the form $\rm \alpha\times [Fe/H] + \beta$ derived in this paper for candidate simple-population GCs and by combining the results of T20 on mass-loss of 1G stars in 46 GCs and those of this paper. We also provide the Pearson rank coefficient, $\rm R_P$, and the r.m.s of the residuals with respect to the best-fit line.}
    \begin{tabular}{l|cccl}
    \hline
    Var. &$\rm \alpha$&$\rm \beta$&$\rm R_P$& scatter\\
    \hline
    &\multicolumn{4}{c}{Simple-population GCs}\\
    \hline
	$\rm \mu$&$0.090\pm0.013$&$0.301\pm0.026$&0.93&$0.019$\\ 
	$\rm M^{HB}$&$0.002\pm0.013$&$0.688\pm0.022$&-0.89&$0.027$\\ 
    \hline
    &\multicolumn{4}{c}{Entire sample}\\
    \hline
	$\rm \mu$&$0.095\pm0.006$&$0.312\pm0.011$&0.89&$0.029$\\ 
	$\rm M^{HB}$&$-0.026\pm0.007$&$0.619\pm0.013$&-0.33&$0.034$\\ 
	\hline
	\hline
	\end{tabular}
	\label{tab:fits}
\end{table}

%%%%%%%%%%%%%%%%%%%%%%%%%%%%%%%%%%%%%%%%%%%%%%%%%%%%%%%%%%%%%%%%%%
\begin{figure*}
    \centering
    \includegraphics[width=1.0\columnwidth]{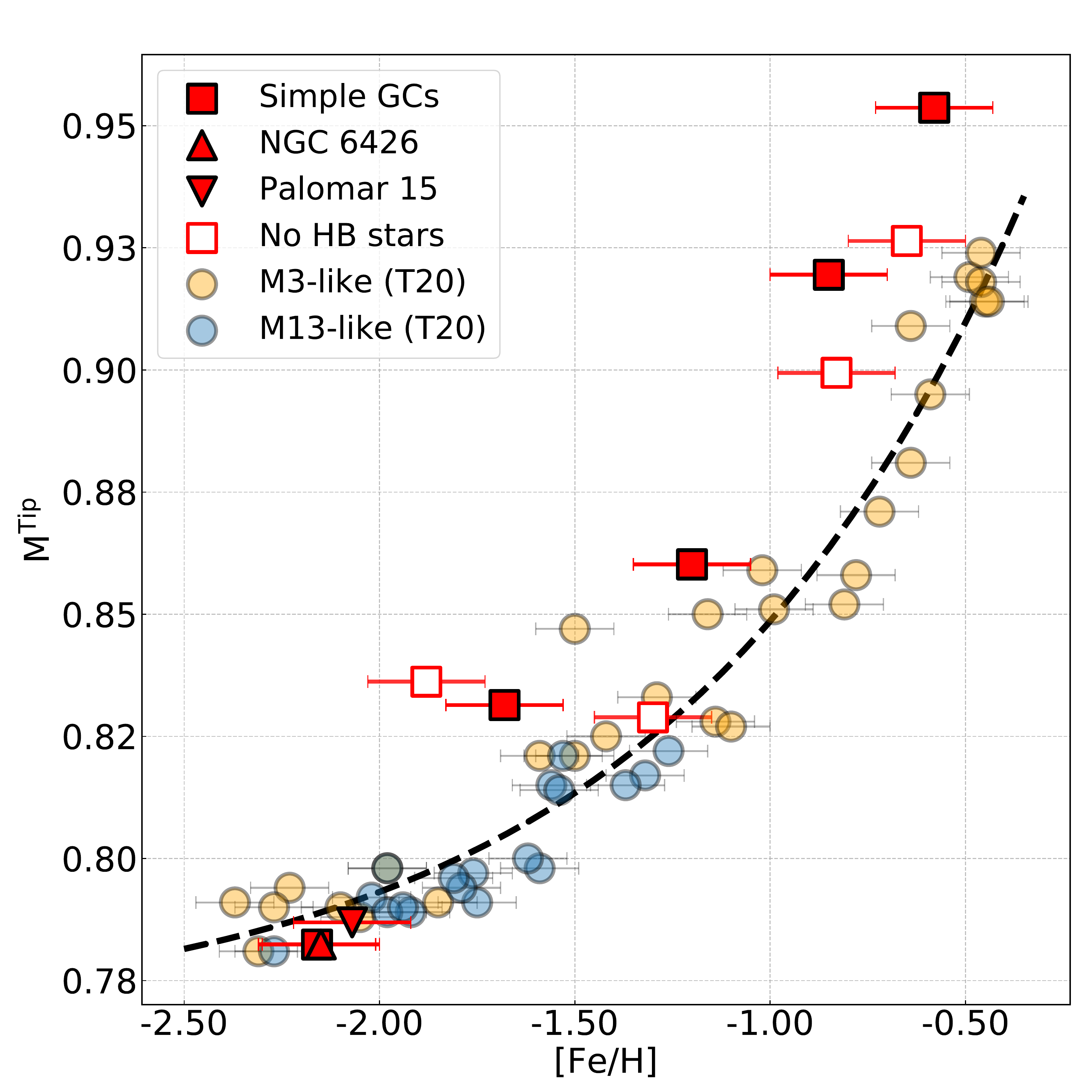}
    \includegraphics[width=\columnwidth]{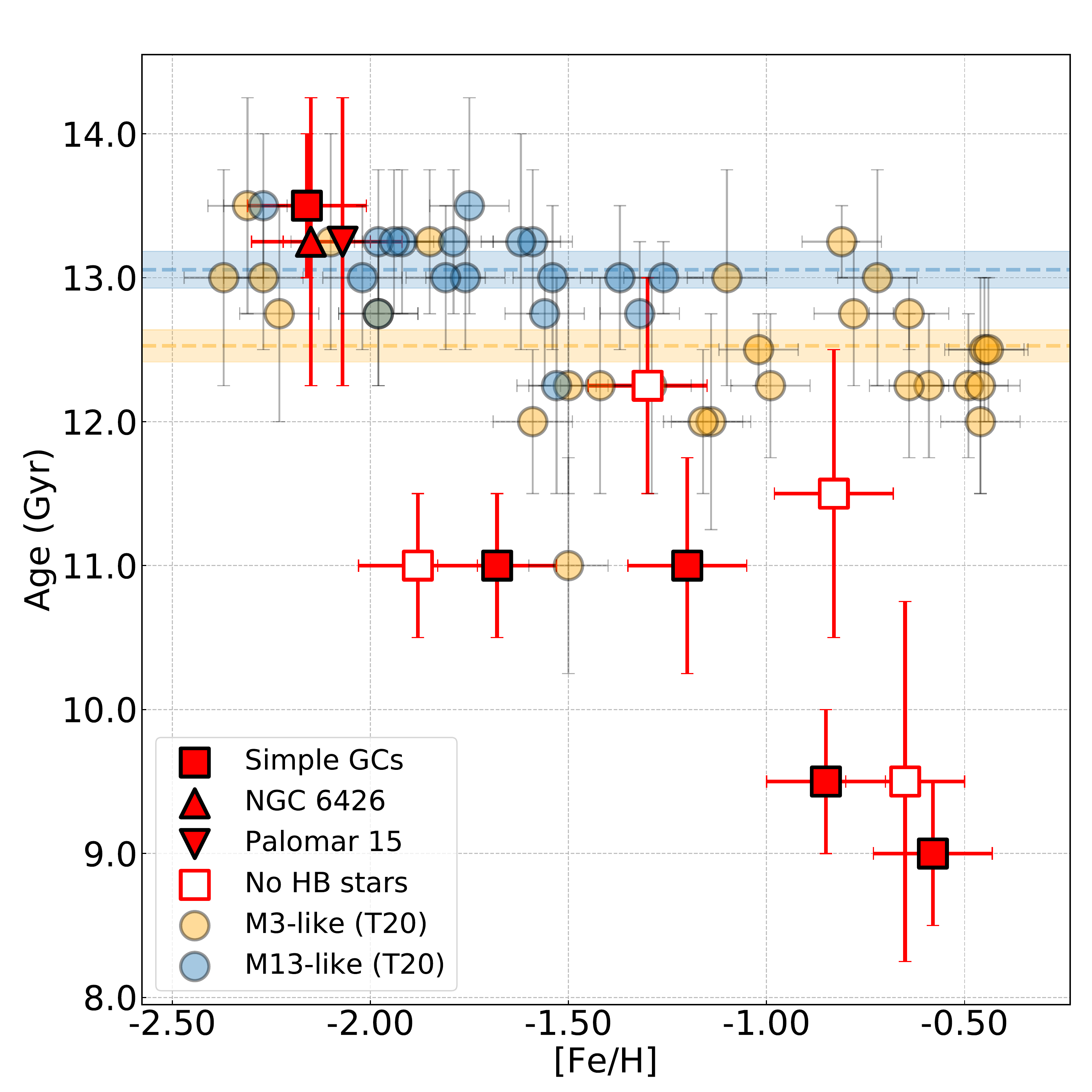}

    \caption{
    Stellar mass at the tip of the RGB ($\rm M^{Tip}$, left panel) and cluster age (right panel) against iron abundance. 
    Clusters with no HB stars are indicated with open squares, while the other symbols are the same as in Figure \ref{pic:mu_feh}. 
     The black dashed line plotted in the right panel is the least-square fit for all T20 clusters, whereas the orange and azure dashed horizontal lines in the right panel indicate the average ages of M\,3 and M\,13-like GCs, respectively. The corresponding $1-\sigma$ age intervals are indicated by the shaded areas. 
    }
    \label{pic:mtip_feh}
\end{figure*}
%%%%%%%%%%%%%%%%%%%%%%%%%%%%%%%%%%%%%%%%%%%%%%%%%%%%%%%%%%%%%%%%%%

To further investigate HB stars in simple-population GCs we show in the left panel of Figure \ref{pic:mtip_feh} the stellar mass at the RGB inferred from the best-fit isochrone against metallicity. In this figure, we also included the studied clusters with no HB stars.
Clearly, candidate-simple population GCs exhibit higher values of $M_{\rm tip}$ than GCs of similar metallicity. 

The large RGB-tip stellar masses are mostly due to the fact that the majority of the simple-population clusters are younger than the bulk of GCs studied by T20.
This is illustrated in the right panel of Figure \ref{pic:mtip_feh} where we show the age-metallicity relation for the clusters studied in this paper (squares) and by T20 (circles).

We confirm previous result by 
 \citet[][and references therein]{dotter_2010,dotter_2011} and \citet[][and references therein]{Leaman_2013} of two main branches of clusters in the age vs.\,[Fe/H] diagram, with simple population GCs populating the younger branch.
 From the comparison between observed GC ages and simulations of Milky-Way formation, Dotter and collaborators suggested that the distinct branches of clusters in the age-metallicity plane may originate from two different phases of Galaxy formation, including a rapid collapse followed by a prolonged accretion \citep[see also][]{kruijssen_2019}. 
 Similarly, based on the integrals of motions by \citet{massari2019a}, \citet{milone2020a} suggested that simple-population GCs may form in dwarf galaxies that have been later accreted by the Milky Way.
  The possibility in-situ and accreted clusters show similar mass loss-metallicity relations further corroborates the evidence of a universal mass-loss law.

\begin{figure}
    \centering
    \includegraphics[width=\columnwidth]{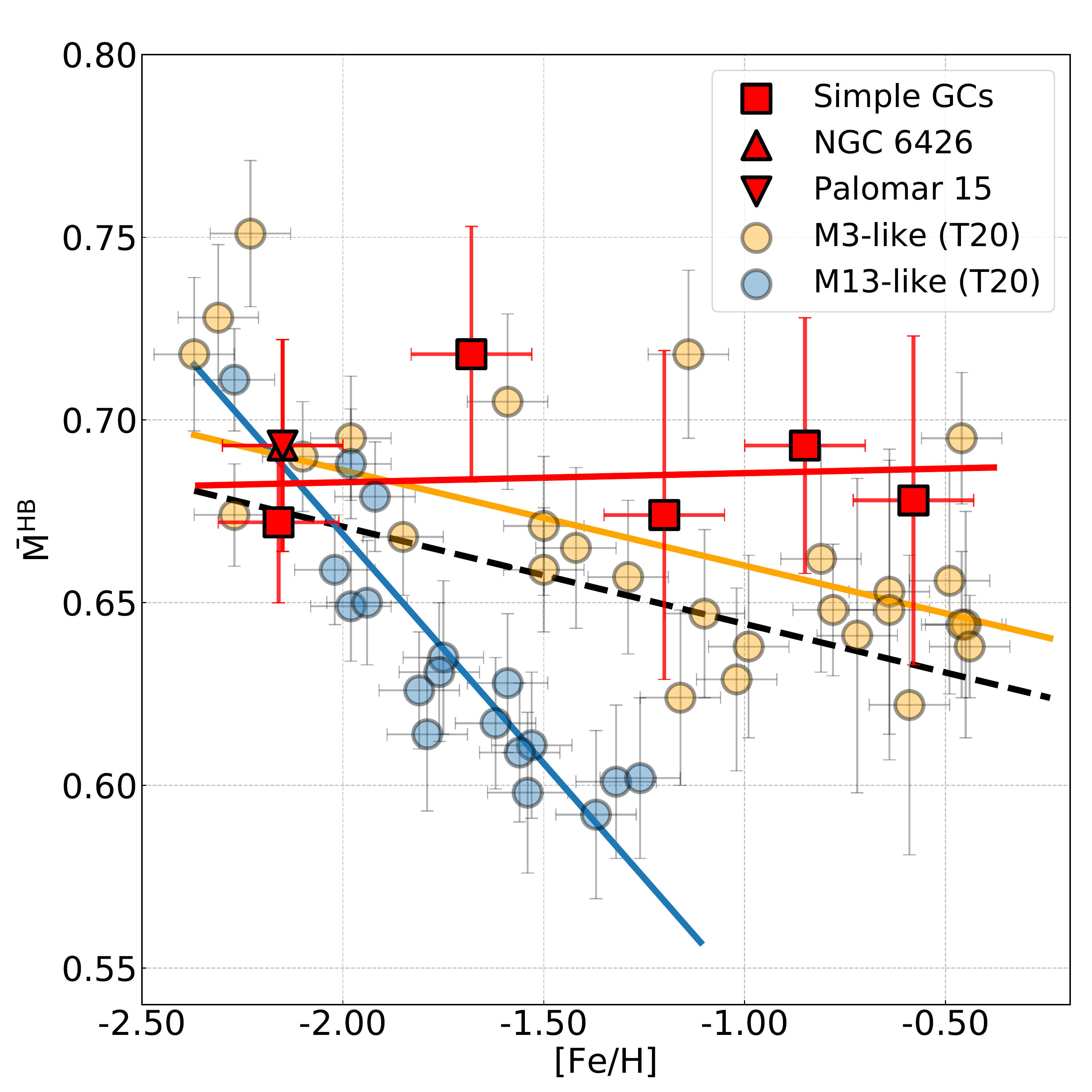}
    \caption{
    Average HB mass ($\rm \bar{M}^{HB}$) as a function of [Fe/H] values for clusters studied in this paper and by T20. 
    The symbols and the colour coding are the same as in Figure \ref{pic:mu_feh}. Red, orange and azure continuous lines are the straight lines that provide the best fit with simple-population candidate, M\,3 like and M\,13-like GCs. The black-dashed line refers to all clusters studied by T20.   
    }
    \label{pic:mhb_feh}
\end{figure}

 Candidate simple-population GCs share similar HB masses as shown in Figure \ref{pic:mhb_feh}, where we plot $\rm M^{HB}$ against [Fe/H]. The HB-mass range of candidate simple-population GCs is comparable with that of 1G stars in M\,3-like GCs and significantly differs from the behaviour of M\,13-like GCs.

The evidence that M\,13-like GCs exhibit different patterns than simple-population star clusters in both the $\mu$ vs\,[Fe/H] and the $\rm M^{HB}$ vs.\,[Fe/H]
 planes indicates that their 1G stars behave differently than the bulk of stars with similar metallicity. As a consequence, in addition to metallicity, some second parameter is responsible for the different mass loss required in their 1G stars. 
 Although M\,13-like GCs are, on average, older than M\,3-like GCs,  age difference alone is not able to account for the different HB shapes. Results of this paper, and from T20, indicate that either mass loss is the second parameter of the HB morphology or 1G stars of all GCs share the same mass loss, but the reddest stars of M\,13-like GCs are enhanced by 0.01-0.03 in helium mass fraction with respect to M\,3-like ones. In this case, as suggested by \citet{dantona_2008}, M\,13-like GCs could have lost all 1G stars and their red HB tails are populated by 2G stars with moderate helium enhancement.

\section{summary and conclusions}
\label{sec:Conc}

We derived high-precision ACS/WFC photometry in the F606W and F814W of seven GCs that are candidates simple stellar populations. We identified probable cluster members by means of stellar proper motions and corrected the photometry for the effects of differential reddening. The resulting CMDs have been used to infer the RGB mass loss by comparing the observed HB stars with appropriate simulated CMDs. The main results can be summarized as follows.

\begin{enumerate}[(i)]
    \item The RGB mass loss in candidate simple-population GCs varies from cluster to cluster and strongly correlates with the cluster metallicity.
    \item The mass-loss vs. [Fe/H] relation is consistent with a similar relation inferred by T20 for 1G stars of 46 GCs.
     We combined the results from this paper and those derived from 1G stars by T20 to derive an improved mass-loss metallicity relation.
          Moreover, our relation matches the values of mass loss inferred by \citet{savino_2019} for dwarf galaxies but is not consistent with the mass-losses values inferred for 2G stars by T20.
         
    \item For a fixed metallicity, the mass losses and the average HB masses of 1G stars in a subsample of GCs with the blue HB alone (M\,13-like GCs) significantly differ from those inferred from simple-population GCs and from 1G stars of the remaining GCs (M\,3-like GCs). 
\end{enumerate}

 These results suggest that the tight correlation between the amount of RGB mass loss and [Fe/H] that we observed both in simple-population GCs and in 1G stars of multiple-population GCs does not depend on the multiple-population phenomenon and is a good candidate as a general property of Populations II stars.
 Moreover, the finding that M\,13-like GCs exhibit different mass-loss vs.\,metallicity relation than simple-population clusters suggests that mass loss is one of the main second parameters that govern the HB morphology of GCs.
    
\section*{Data Availability}
The images analysed in this work are publicly available one the Space Telescope Science Institute \textit{HST} web page (https://www.stsci.edu/hst) and on the Mikulski Archive for Space Telescopes portal (MAST, https://archive.stsci.edu/). The photometric and astrometric catalogs underlying this article will be released on the Galfor project web page 
 (http://progetti.dfa.unipd.it/GALFOR/) and on the CDS site. The models obtained via the ATON 2.0 code are not yet available to the public but they can be provided upon request.
    
\section*{Acknowledgements}
This work has received funding from the European Research Council (ERC) under the European Union's Horizon 2020 research innovation programme (Grant Agreement ERC-StG 2016, No 716082 'GALFOR', PI: Milone, http://progetti.dfa.unipd.it/GALFOR). MT, APM and ED acknowledge support from MIUR through the FARE project R164RM93XW SEMPLICE (PI: Milone). MT, APM and ED have been supported by MIUR under PRIN program 2017Z2HSMF (PI: Bedin). EV acknowledges support from NSF grant AST-2009193.

%%%%%%%%%%%%%%%%%%%%%%%%%%%%%%%%%%%%%%%%%%%%%%%%%%

%%%%%%%%%%%%%%%%%%%% REFERENCES %%%%%%%%%%%%%%%%%%

% The best way to enter references is to use BibTeX:

\bibliographystyle{mnras}
\bibliography{HBsimple} 

%%%%%%%%%%%%%%%%%%%%%%%%%%%%%%%%%%%%%%%%%%%%%%%%%%

% Don't change these lines
\bsp    % typesetting comment
\label{lastpage}
\end{document}